Trusses and Trapezes: Easily-Interpreted Communities in Social Networks

Jonathan Cohen
Jonathan.Cohen@jhuapl.edu
Johns Hopkins University Applied Physics Laboratory
11100 Johns Hopkins Road, Laurel, Maryland 20723

# Abstract

The truss, a relaxation of the clique based on triangles, serves to identify clusters of actors in a way that is easy to interpret and is computationally attractive. This paper introduces the 4-cycle-based relative to the truss, called the trapeze, presents a weighted extension of both the truss and trapeze, and offers the refinements of strong trusses and trapezes and summit trusses and trapezes. Use of trapezes permits the application to bipartite graphs, while the weighted versions permit variation of support due to natural edge weights without increasing computational complexity. Finally, strong and summit versions make for easy determination of communities across graphs of varying density. Each of these structures offers guaranteed computation in polynomial time, is motivated by a natural observation of social cohesion, and is related nicely to other standard structures.

# Keywords

# Introduction

Graphs are frequently used to capture and study patterns of ties between social actors. Actors are represented by vertices (or nodes); edges (or links) between the vertices represent relationships among the corresponding actors. The simple pictorial nature of graphs goes a long way toward human comprehension of the network of relationships, and the structural nature of graphs provides easy automated analysis.

Many of the operations performed on social network graphs seek to identify cohesive communities of actors, represented by subgraphs within the larger network, for further analysis. The communities are interesting in their own right and help to illuminate the roles of actors on their frontiers.

In 2001, Moody observed that while most of the methods of social network analysis were developed for necessarily-small graphs, social analysis was attempting to address networks of rapidly-increasing size, a trend that has exploded today. Accordingly, it is important to identify methods that are not only effective, but efficient.

The most obvious cohesive subgraph, introduced to social network analysis by Luce and Perry (1949), is the clique — a subgraph in which every vertex is adjacent to every other vertex. Usually, the definition further specifies that the clique be maximal, so that no other vertices may be added to the subgraph without it ceasing to be a clique. The clique turns out to be of disappointing utility because it is both too rare and too common: cliques of only a few members are frequently too numerous to be helpful, and larger cliques are too limiting to be expected even among tightly-knit actors, particularly in light of incomplete information. In addition, the automated finding of cliques is intractable, that is, the work scales worse than any polynomial of the problem size, making the enumeration of cliques impractical for even moderate data sizes (Bron and Kerbosch, 1973). Moreover, cliques prove to be difficult to interpret because they are likely to be horribly entangled. Consider, for example, a social network consisting a single clique of 20 actors. If one were to remove a few edges from such a graph, the result would be several cliques of 19 actors, with most of the memberships in common across the cliques. Thus, a group of actors that should still be considered a single cluster becomes a clutter of intersecting cliques. The cliques in a more general and realistic graph are even more difficult to use.[1]

---

[1] One can record statistics of joint membership in cliques and use them as a basis for clustering, but this merely defers clustering.

One may choose to generalize and relax the clique to avoid these problems. This process can take several different roads, depending on which properties are maintained and which are loosened.

One property of a clique is that each member is a distance 1 away from every other member, that is, one can get between any pair of actors by traversing only a single edge. The so-called *n*-clique[2] (Luce, 1950) is a generalization of the clique, and consists of a group of vertices such that each pair of members is separated by no more than a distance *n*. With this nomenclature, a clique is a 1-clique. The *n*-clique suffers from quite a few problems: For *n* > 2, the group is likely to be too diffuse to be of interest. Also, the problem of enumerating *n*-cliques is no better than enumerating cliques. Finally, it may happen that intermediaries responsible for proximity of an *n*-clique's members may not be members of the *n*-clique itself (Alba, 1973).

Two attempts to address problems of *n*-cliques are *n*-clans and *n*-clubs (Alba, 1973 and Mokken, 1979). *n*-clans are *n*-cliques whose diameter[3] is no bigger than *n*. An *n*-club is a maximal subgraph of diameter *n*. In both cases, the diameter is measured strictly within the subgraph. These modifications do not remove the problems of enumeration (there are still too many), interpretation, or of computational intractability.

Another property of a clique is that each member is adjacent to all clique members save itself. A generalization in this respect is the *k*-plex (Seidman and Foster, 1978). A *k*-plex is a maximal subgraph in which each vertex is adjacent to all but at most *k* of the members; a clique is a 1-plex. Again, enumeration and intractability are no better than for cliques.

The simple *k*-core, introduced by Seidman (1983), is a maximal (single-component) subgraph in which each member is adjacent to at least *k* other members. It, too, is a generalization of the clique: a clique is a *k*-core with *k*+1 members. Unlike the other clique generalizations above, the *k*-core can be computed in polynomial time, and does not pose an enumeration problem. The disadvantage of *k*-cores is that they are too promiscuous: rather than being sets of high cohesion, Seidman characterizes *k*-cores as "seedbeds, within which cohesive subsets can precipitate out." The *k*-core does narrow the search for cohesive subgroups, but at the expense of including much else.

---

[2] An unfortunate name, since *n*-clique usually means a clique of *n* actors, and will have that meaning throughout the remainder of this communication.

[3] Diameter, when an integer as opposed to a path, is the largest distance between two vertices in the graph, where distance between two vertices is defined as the minimum number of edges that must be traversed to go from one vertex to the other.

In addition to the methods cited above, many other approaches to finding graph communities have been taken.

Cohesive groupings may be defined on the basis of comparing internal connections to external ones; vertices are added to sets to maximize intragroup ties while minimizing intergroup ties. The popular method of maximizing this internal to external "modularity," introduced by Clauset, *et al.* (2004), is quite fast and easily interpreted. Improving on this, Blondel, *et al.* (2008), offer a modularity-optimization approach that allows reexamination of early assignments. But maximizing modularity has several unfortunate characteristics (Fortunato, *et al.*, 2007, Lancichinetti, *et al.*, 2011a). Earlier related work on LS-sets, borrowed from circuit partitioning for board layout (Luccio and Sami, 1969), are so special as to be extremely rare in social networks and are expensive (but polynomial) to calculate (Borgatti, *et al.*, 1990). A relaxation of LS-sets, $\lambda$-sets, still requires quartic work (Borgatti, *et al.*, 1990).

Label passing (Raghavan, *et al.*, 2007) is an example of a dynamical systems method in which community structure is determined by collecting information from adjacent vertices.

Spectral methods, based on eigenvectors of matrices associated with the graph, detect communities through a hierarchy of clusters by repeated bipartition, each informed by blocks induced by thresholding elements of eigenvectors. Also in the vein of divisive clustering methods, Girvan and Newman (2003) proposed partitioning by repeated removal of the edge of largest betweenness, a rather expensive method only suitable for gross segmentation of the graph.

These approaches, and many more, appear in the review of graph community detection algorithms offered by Fortunato (2010).

In addition to methods that force each vertex into exactly one community, many recent approaches have sought ways to allow actors to belong to multiple communities, that is, to allow communities to overlap:

The OSLOM method (Lancichinetti, *et al.*, 2011b) seeks to accrete communities that exhibit statistically-significant connectivity against a null model of random edge assignment that respects nodes degrees. Like the method of Blondel, *et al.* (2008), an important feature of this method is that it continually reevaluates existing membership and is thereby able to overcome early false moves. OSLOM allows vertex membership in multiple communities and accommodates directionality and weights. Also permitting multiple membership, clique percolation (Palla, *et al.*, 2005) constructs communities by finding small cliques and joining adjacent ones (those sharing all but one vertex). Basing the construction on finding cliques makes the process rather expensive, but is mitigated somewhat by restricting the cliques to small size.

A review and comparison of popular overlapping community detection methods is offered by Xie *et al.* (2013). For a general introduction to social network analysis, the interested reader may consult the book by McCulloh, *et al*. (2013). Wasserman and Faust (1994) offer a more in-depth and extensive presentation. Both discuss a variety of cohesive subgraphs and their application.

One would like to employ something between the expensive-to-find and overly-numerous groupings provided by cliques, *n*-cliques, *n*-clans, *n*-clubs, and *k*-plexes on the one hand, and the inexpensive, few-in-number, but overly-generous *k*-cores on the other.

Toward that end, in the early 2000s the author defined *trusses* and *trapezes* and some of their extensions, and implemented them in a general-purpose graph computer program called Renoir, which saw distribution within the federal government and enjoyed commercial licensing beginning in 2001 and continuing to the present.[4] In 2008 he published some of that work as background information to a paper on graph computing using MapReduce (Cohen, 2008). Despite the obscure disclosure, that background paper has been cited by several authors (Redmond, *et al.*, 2011, Wang and Cheng, 2012, Huang, *et al.*, 2014, Chen, *et al.*, 2014, Sariyuce, *et al.*, 2015, Malliaros, 2015, and, Zheng, *et al.*, 2017.). The truss was also rediscovered in another setting (Zhang and Parthasarathy, 2012). The current communication provides a more complete exposition of that work and its context, introduces the trapeze, and offers interesting generalizations and specializations, including the ability to accommodate edge weights. The truss, trapeze, strong truss, and strong trapeze were implemented in Renoir; the weighted and summit variations were not.

The truss is another generalization of the clique, and is based on 3-cycles, that is, triangles, in the graph. In a similar manner, the *trapeze* is built on 4-cycles (rectangles), and is particularly useful for bipartite graphs, which necessarily lack triangles.

The paper motivates the definition of the truss, offers properties of it, then introduces its cousin, the trapeze. The *strong* versions of both are then presented, which identify particularly tight groups. After presenting some examples, the paper goes on to present new definitions that had not been present in Renoir, namely weighted trusses and trapezes and summit trusses and trapezes.

---

[4] NSA Technology Transfer Program, DHS Digital Library, https://www.hsdl.org/?view&did=738432

# The Truss

White and Harary (2001) argue that what has been missing from many of the definitions of cohesive groups is the strength from having multiple paths joining actors for a given social relation.[5] It is this point of departure that leads to the *truss*. In particular, one can make this observation about social structures: if two actors are strongly tied, it is likely that they also share ties to others. Turning this around, we may say that an observed edge between two actors is more likely to be significant if they have common neighbors. Based on this observation, we shall create something called a *k*-truss, in which a tie between *A* and *B* is considered legitimate only if "supported" by at least *k*–2 other actors who are each tied to *A* and to *B*.[6] But we will also demand this: that the edges that offer such support for the legitimacy of an edge must, in turn, be supported in a similar manner, and so on. The word "support" here is used to suggest that a bigger number buttresses the hypothesis that an edge is not random.

The truss may be considered yet another relaxation of the clique. One may view the clique of order *k* as a subgraph in which each edge joins two vertices that have at least *k*–2 common neighbors. Stated another way, each edge is supported by *k*–2 pairs of edges making a triangle with the original edge. Here is the formal definition:

**Definition:** A *k*-truss is a non-trivial, one-component subgraph such that each edge is supported by at least *k*–2 pairs of edges, within the subgraph, making a triangle with that edge. (Non-trivial here excludes an isolated vertex as a truss.)

In the absence of words to the contrary, the term "*k*-truss" will mean "maximal *k*-truss" — one that is not a proper subgraph of another *k*-truss for fixed *k*.

The maximal 2-trusses are the components of the graph. *k*-trusses for *k* less than 2 are undefined.

---

[5] White and Harary go on to suggest cohesive groups identified by vertex connectivity, a criterion that is stronger, and more expensive, than the truss.

[6] The author regrets an aspect of this definition, preferring now to say that a *k*-truss has edge support of *k*, rather than *k*–2. But his original definition was adopted by others, and he feels compelled to remain consistent here.

**Observation:** A clique of order $k$ is a $k$-truss.

**Proof:** Let $G$ be a $k$-clique and let $(a,b) \in E(G)$. Since $|V(G)| = k$, we have distinct vertices $v_1, v_2, \ldots, v_{k-2} \in V(G) - \{a,b\}$. Since $G$ is clique, for $i = 1, 2, \ldots, k-2$, $(a, v_i), (v_i, b) \in E(G)$. Therefore, $(a,b)$ is in $k-2$ triangles.

An example of trusses is offered by Figure 1. The graph is of frequent associations between 62 dolphins in a community living off Doubtful Sound, New Zealand, taken from Lusseau, *et al.* (2003). The graph contains a single (maximal) 3-truss, shown in Figure 1(b). It has four 4-trusses, only one of which is a clique. Figure 1(d) shows there are two 5-trusses, one of which is a clique of order 5, the other a clique of order 6, missing one edge.

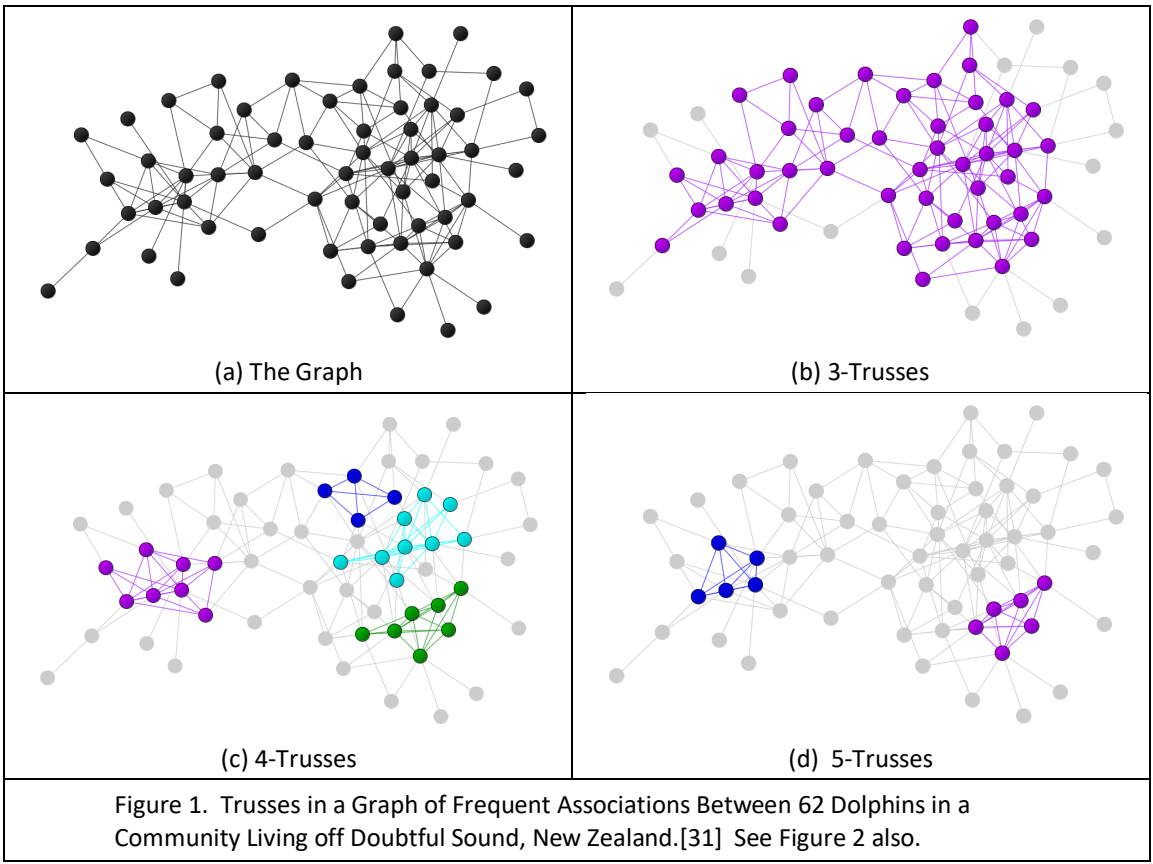

Figure 1. Trusses in a Graph of Frequent Associations Between 62 Dolphins in a Community Living off Doubtful Sound, New Zealand.[31] See Figure 2 also.

Figure 2 shows the right-most 5-truss from Figure 1(d). Clearly, it identifies a tight group of dolphins, despite failing to be a clique.

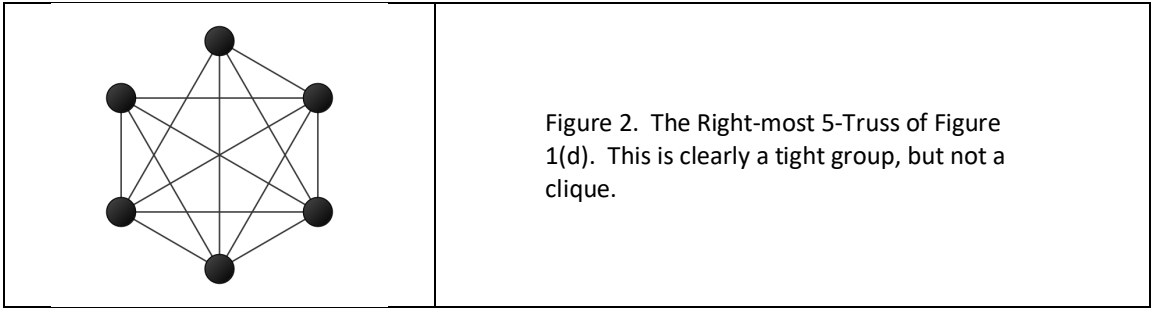

Figure 2. The Right-most 5-Truss of Figure 1(d). This is clearly a tight group, but not a clique.

Another example is offered by Figure 3, in which the nodes represent political books, and edges between them indicate frequent co-purchases from Amazon.com. This data comes from Krebs (2004). Each book was examined (by Krebs) and judged to be liberal, conservative, or neutral, and given a corresponding icon of an upward-pointing triangle, a downward-pointing triangle, or a rhombus, respectively.

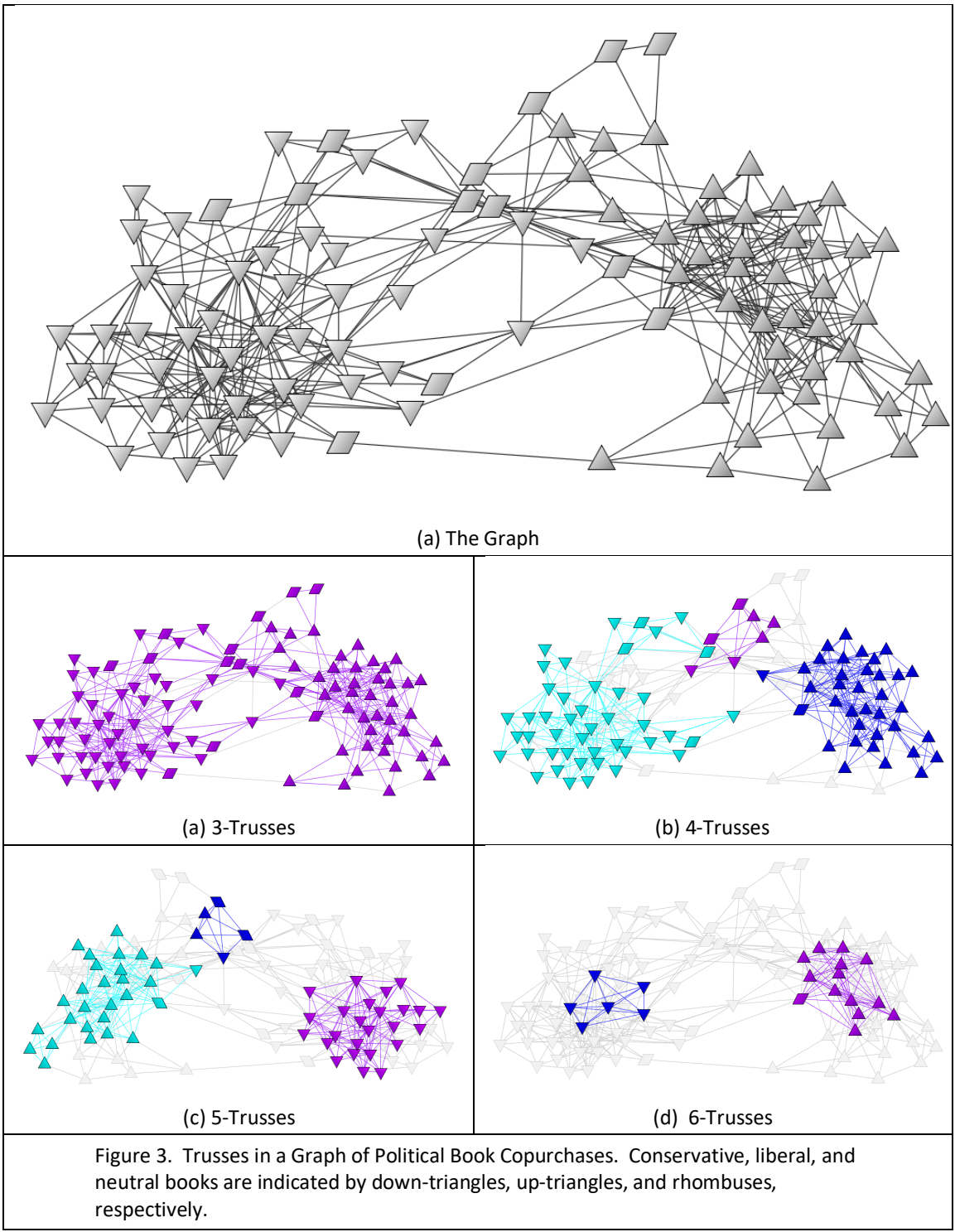

Figure 3. Trusses in a Graph of Political Book Copurchases. Conservative, liberal, and neutral books are indicated by down-triangles, up-triangles, and rhombuses, respectively.

The trusses seem to quickly identify political camps, including one apparently neutral one. The left-most 6-truss is a clique, the right-most one is far more complicated and defies further subdivision. Curiously, the right-most truss of Figure 3(d), the core of the liberal books, contains one book labeled neutral. Reviews for this book on Amazon.com show disagreement as to whether the book is liberal or neutral (Amazon.com).

Trusses are easy to interpret, according to this observation:

**Observation:** Maximal $k$-trusses (for any fixed $k$) do not intersect.

Trusses are nested:

**Observation:** For $k > 2$, each $k$-truss is inside of a $k$–1 truss.

Trusses occur inside of $k$-cores:

**Observation:** Each $k$-truss of $G$ is a subgraph of a $(k–1)$-core of $G$.

Sometimes we have a (maximal) truss that can be viewed as smaller non-maximal trusses joined by a small number of nodes. When the set of edges in a $k$-truss can be divided such that edges in each subset don't benefit from edges in another subset for their support, the $k$-truss can be separated into *strong trusses*.

**Definition:** Two edges $a$ and $b$ are *triangle-adjacent*, denoted $a\Delta b$, if $a$ and $b$ share a triangle.

**Definition:** Two edges $a$ and $b$ are *triangle-connected* if $a\Delta b$ or if there exist edges $l_1, l_2, \ldots, l_n$, such that $a\Delta l_1$, $l_n\Delta b$, and $l_i\Delta l_{i+1}, i = 1, \ldots, n-1$.

**Definition:** A strong $k$-truss is a (perhaps non-maximal) $k$-truss such that each pair of its edges are triangle-connected.[7]

Unless stated otherwise, the term strong truss will mean a maximal strong truss.

We return to the example of Figure 1, but this time we look at strong trusses contained in the 3-truss of Figure 1(b), shown in Figure 5. The figure shows that the single truss can be viewed as the union of four smaller groups, joined by a few

---

[7] When maximal, Huang, *et al.* (2014), call this a *truss community*. Their definition was based on a weaker definition of a truss, which did not require a truss to be a single component subgraph.

individuals.

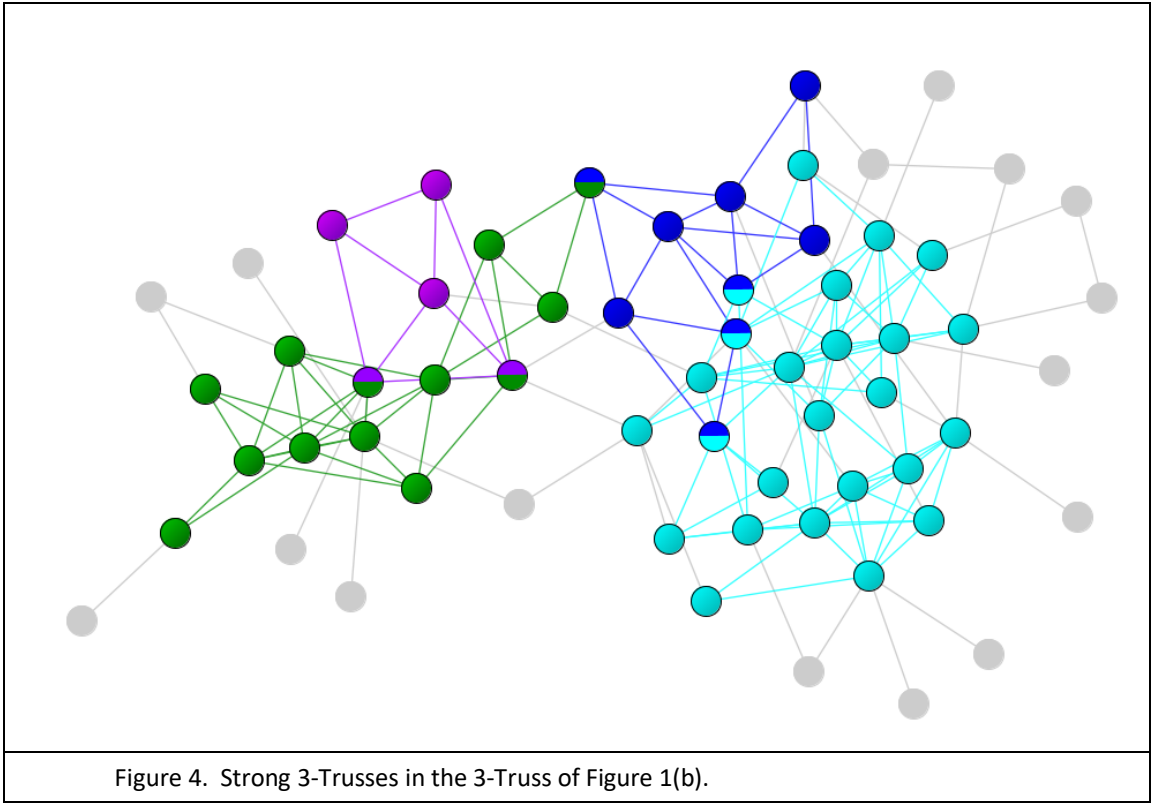

Figure 4.  Strong 3-Trusses in the 3-Truss of Figure 1(b).

The point of considering strong trusses is that they reveal further structure and identify actors that form (perhaps crucial) bridges between social groups.

# The Trapeze

In the event that the graph under examination is bipartite, no triangles will be present, but one can take a similar approach with rectangles, that is, with 4-cycles. Here, we will take the significance of an edge as being demonstrated by the support of other edges such that it is part of some minimum number of rectangles.

**Definition:** A *k*-trapeze is a non-trivial, one-component subgraph such that each edge is supported by at least *k* distinct sets of 3 edges, within the subgraph, making a 4-cycle with that edge. (Non-trivial here excludes an isolated vertex as a trapeze.)

In the absence of words to the contrary, the term trapeze will mean maximal trapeze.

Like trusses, trapezes are easy to interpret, according to this observation:

**Observation:** Maximal $k$-trapezes (for any fixed $k$) do not intersect.

Trapezes are nested:

**Observation:** For $k > 1$, each $k$-trapeze is inside of a $k-1$ trapeze.

Like the strong truss, we can define a strong trapeze.

**Definition:** Two edges $a$ and $b$ are *rectangle-adjacent* denoted $a\blacksquare b$, if $a$ and $b$ share a rectangle, that is, a 4-cycle.

**Definition:** Two edges $a$ and $b$ are *rectangle-connected* if $a\blacksquare b$ or if there exist edges $l_1, l_2, \ldots, l_n$, such that $a\blacksquare l_1$, $l_n\blacksquare b$, and $l_i\blacksquare l_{i+1}, i = 1, \ldots, n-1$.

**Definition:** A strong $k$-trapeze is a (perhaps non-maximal) $k$-trapeze such that each pair of its edges are rectangle-connected.

Unless stated otherwise, the term "trapeze" will mean "strong trapeze."

Here is an example: DonorsChoose.org offers data describing all of the donations to teachers through their organization (DonorsChoose.org). Each transaction gives a hash of the donor and of the (recipient) project, among other things. From this, a donor-project graph was constructed, having 1M vertices and 1.6M ties leading to 1.3M edges. One would hope that this bipartite graph would reveal giving communities joining actors of similar interests and projects of similar intent. This graph contained 13k strong 1-trapezes (encompassing about a quarter of the graph's edges), 4.7k strong 2-trapezes, 1.2k strong 4-trapezes, 364 strong 8-trapezes, 140 strong 16-trapezes, 43 strong 32-trapezes, 17 strong 64-trapezes, 7 strong 128-trapezes, 2 strong 256-trapezes, 2 strong 512-trapezes, and one strong 1024 trapeze. (Values of support between these values also existed, of course.)

Figure 5 shows one of the strong 4-trapezes. Donors are indicated by a person icon, the recipient projects as rounded rectangles. In this subgraph there is clearly much commonality among donors funding projects and among projects receiving funding.

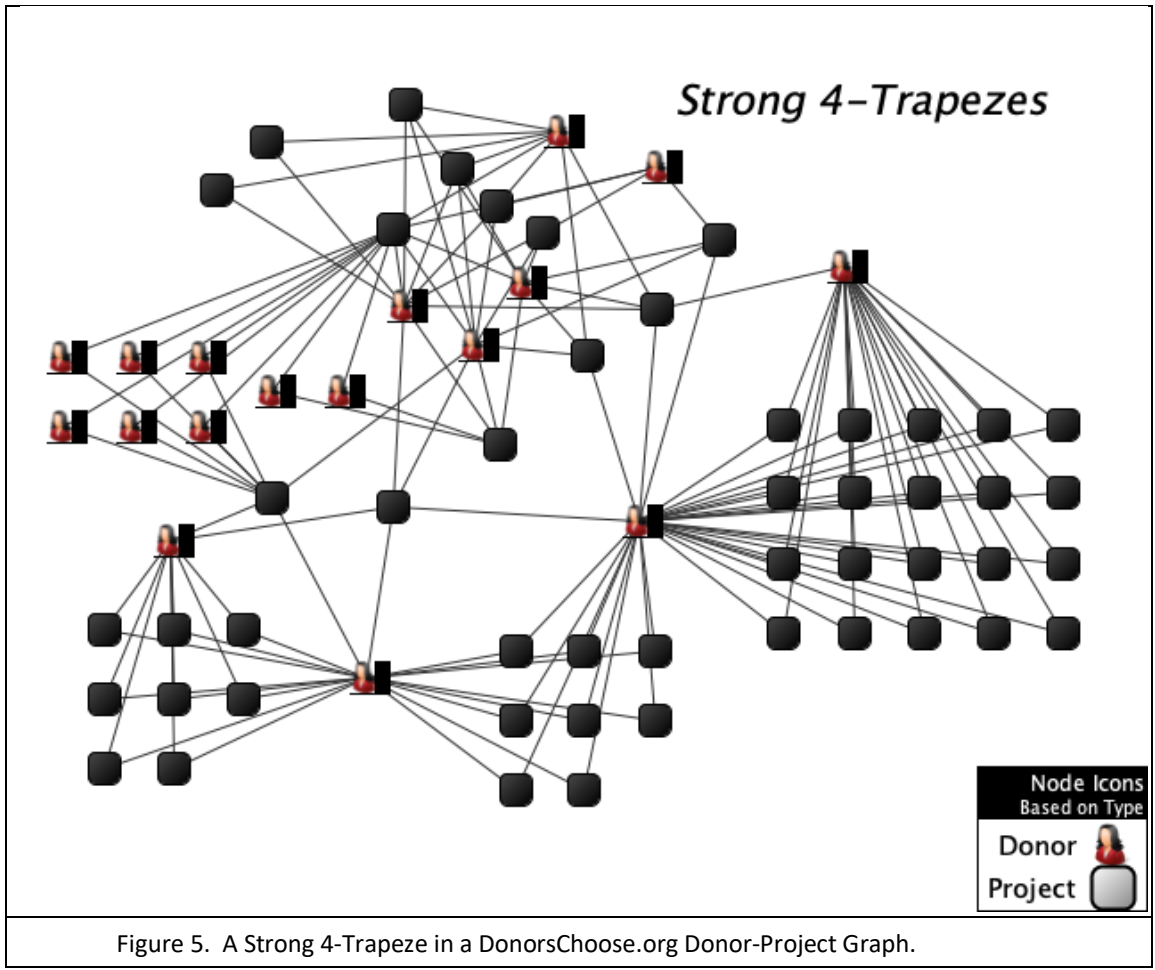
Figure 5.  A Strong 4-Trapeze in a DonorsChoose.org Donor-Project Graph.

When we tighten the support requirement to 8, the strong trapeze of Figure 5 is seen to contain three tighter groups, as shown in Figure 6.  Trapeze membership is indicated by color; the black elements do not belong to any 8-trapeze.  Note that one donor and one project belong to two strong 8-trapezes.

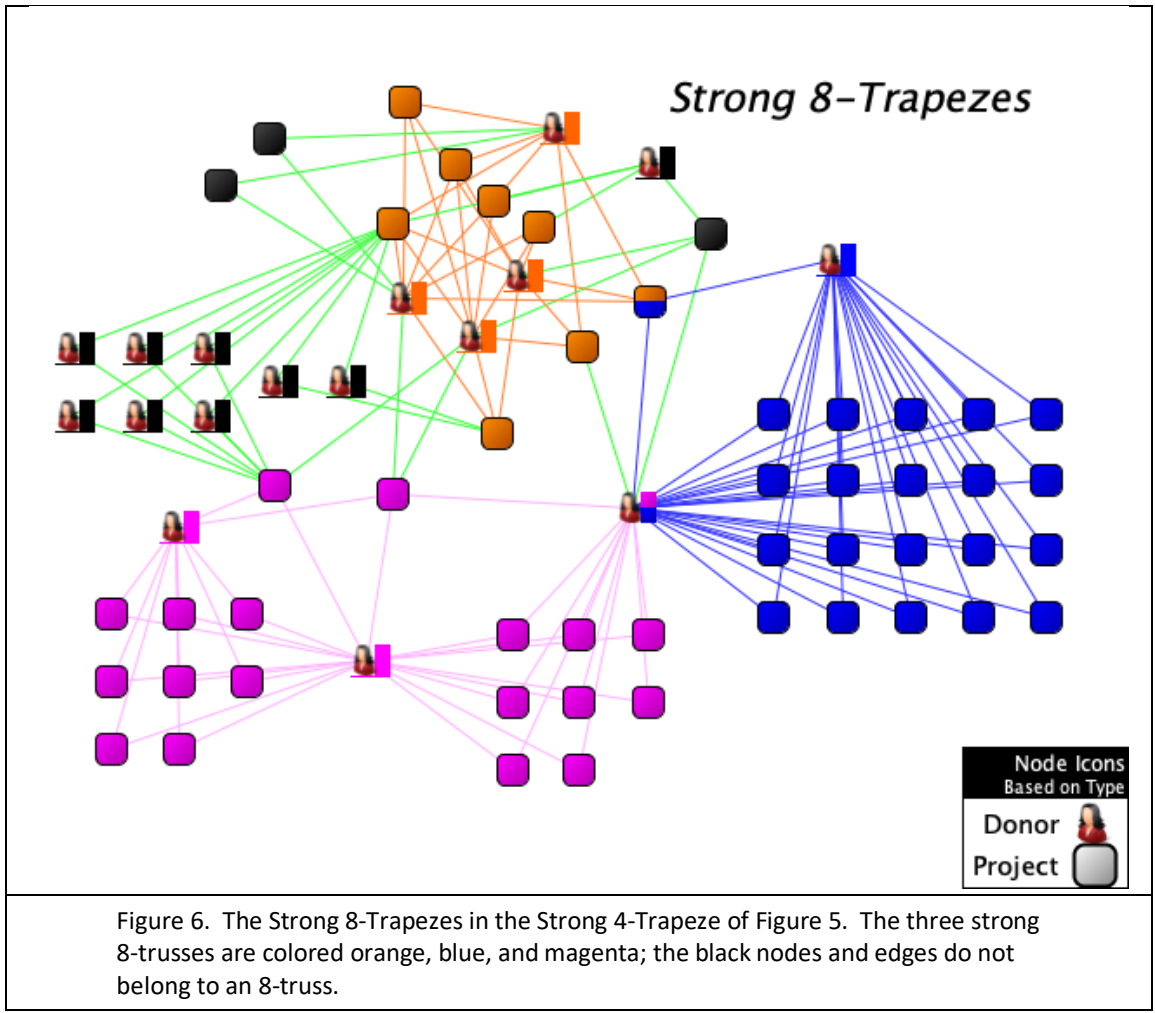

Figure 6. The Strong 8-Trapezes in the Strong 4-Trapeze of Figure 5. The three strong 8-trusses are colored orange, blue, and magenta; the black nodes and edges do not belong to an 8-truss.

The existence of trapezes and of trusses are not independent:

**Observation:** For $k > 3$, each $k$-truss is a $(k-2)(k-3)$-trapeze, though the trapeze need not be maximal.

**Proof:** Choose $k > 3$ and let $G$ be a $k$-truss. Pick arbitrary edge $(a, b) \in E(G)$. Since $G$ is a $k$-truss, there exist distinct vertices $v_1, v_2, \ldots, v_{k-2} \in V(G) - \{a, b\}$ with $(a, v_i), (v_i, b) \in E(G)$. Pick $i \in \{1, 2, \ldots, k-2\}$. Again, by $G$ being a truss, there exist distinct vertices $u_1, u_2, \ldots, u_{k-2} \in V(G) - \{a, v_i\}$ with $(a, u_j), (u_j, v_i) \in E(G)$, $j = 1, 2, \ldots, k-2$. One of $\{u_j\}$ is $b$; label it $u_{k-2}$, so that the cycle $C_{ij} = \big((a, u_j), (u_j, v_i), (v_i, b), (b, a)\big)$ is not degenerate for $j \in \{1, 2, \ldots, k-3\}$. Moreover, each $C_{ij}, i \in \{1, 2, \ldots, k-2\}, j \in \{1, 2, \ldots, k-3\}$ is distinct, so $G$ is a $(k-2)(k-3)$-trapeze.

This means that finding trapezes is more general, and will find trusses (except that it misses the 3-trusses), as well as other structures. For example, the graph of Figure 7 is a single 2-trapeze that contains two 4-trusses and also contains the structure that joins the trusses. It may also be viewed as three strong 2-trapezes, as indicated by the colors.

Finally, we observe that trapezes, like trusses, are relaxations of cliques:

**Observation:** For $k > 3$, each $k$-clique is a $(k-2)(k-3)$ trapeze.

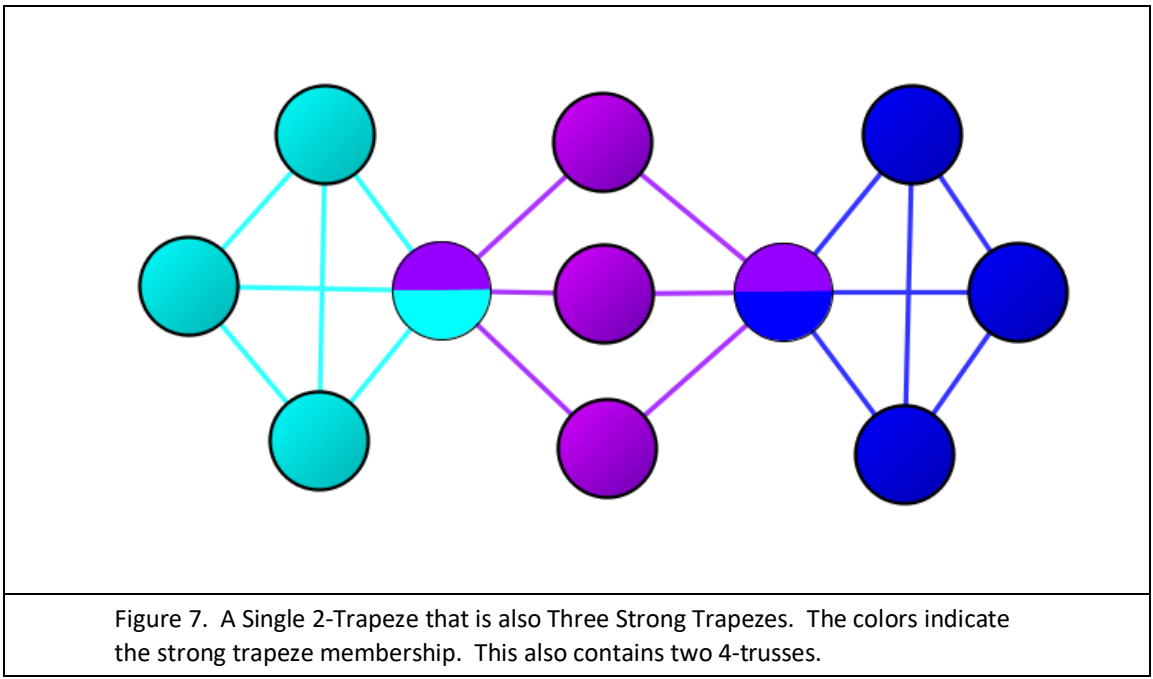

Figure 7. A Single 2-Trapeze that is also Three Strong Trapezes. The colors indicate the strong trapeze membership. This also contains two 4-trusses.

# Weighted Trusses

It often happens that one would like to give some edges more weight than others. Factors that might drive weighting include strength of ties, recency of ties, certainty of information, and value of nodes joined by an edge. In particular, it would be useful to have the concept of support generalized to reflect those weights, so that "stronger" triangles confer more support than weaker ones. Moreover, one would like to make such a generalization without sacrificing the attractive computational complexity enjoyed by unweighted versions of trusses and strong trusses.

It turns out that one can make such a generalization, with some reasonable caveats: First, we will consider only positive weights, excluding the possibility that adding ties can *increase* distance between actors that they join. Secondly, we will note that it is hard to conceive of a realistic situation in which distinguishing between a large number of weights is warranted. Accordingly, we will construct triangle weights that take on integer values over some compact range.

For each edge $e$, define a weight $w(e)$. We will then give each triangle $\Delta$ a weight $W_\Delta \in \mathbb{N}$ based on the weights of its constituent edges.

The function $W_\Delta$ is chosen to be invariant to permutations of its edges and be monotonically increasing in each of its edge's weights. Some obvious choices:

harmonic mean

$$W_\Delta^{\text{harm}} = \left\lfloor \alpha \left( \sum_{e \in \Delta} w^{-1}(e) \right)^{-1} \right\rfloor$$

and minimum

$$W_\Delta^{\min} = \left\lfloor \alpha \min_{e \in \Delta} w(e) \right\rfloor,$$

where $\lfloor x \rfloor$ denotes the integer portion of $x$ and where $\alpha$ is a scaling constant.

With these weights in hand, each edge $e$ enjoys a weighted support of

$$S_w(e) = \sum_{\Delta \ni e \in \Delta} W_\Delta,$$

that is, the weighted support for $e$ is the sum, over all triangles that contain $e$, of the weights of those triangles. Since the weights are restricted to positive integers, each actual triangle appears to confer some number of effective triangles to the support of an edge, that number depending on the weights of the edges making up the triangle.

Figure 8 offers an example, in which the 2012-2013 term of the United States Supreme Court 5-4 decisions are considered (DeSilver, 2013). For each pair of justices, a joining edge carries a weight equal to the percentage (an integer between 0 and 100) of times that the justices agreed. Triangles were given the minimum weight of the constituent edges, so that each triangle conferred a weight equal to the the minimum percentage of agreement. The two strong $k$-trusses for $k \in \{69, 70, 71, \ldots, 130\}$ are shown in the figure, showing the pivotal role of Justice Kennedy.

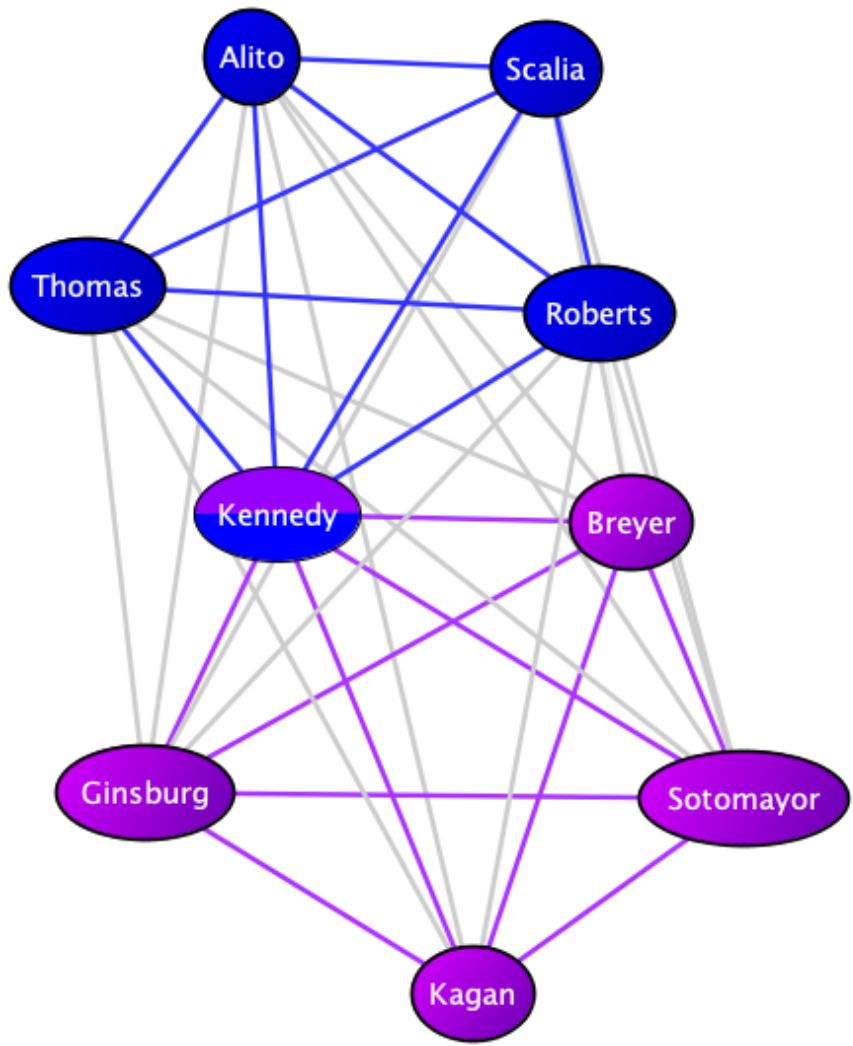

Figure 8. Weighted Strong Trusses in Graph of Agreements Between United States Supreme Court Justices (2012-2013). Trusses are indicated by colors; light green edges are not in the trusses. Edges were given a weight equal to the percentage of time that the judges agreed in 5-4 decisions. Triangle weight was minimum of edge weights taken as integers in {0, 1, …, 100 }. The strong trusses shown exist for (weighted) support 67 through 128.

Before going on, we should note that another type of weighted truss was defined by Zheng, *et al* (2017), in which they define the weight of a truss as the smallest weight of an edge in the truss, and offer methods to determine, for a specified value of $k$ and a specified integer $r$, the $r$ highest-weight $k$-trusses. Their model does not allow the weights to confer different supports.

# Summit Trusses

One annoying facet of the parameter-based clustering approaches is that one often has to try many values of said parameter on each new set of data in the hopes that some value will produce a satisfying result. In the case of trusses, that parameter specifies the support threshold. Moreover, if the parameter governs the tightness of the clusters (as is the case here), a given parameter value may produce acceptable results in one area of the graph and disappointing results in another, with the optimum value for each area largely a function of local density. As there is no global optimum, it would be advantageous to have a value that "floats", effectively demanding higher cohesiveness in denser areas of the graph.

An example is pictured in Figure 9, showing a small portion of a larger Facebook friendship graph collected by McAuley and Leskovec (2012) and available from SNAP.[8] The figure shows strong trusses at various levels. If we wish to see clusters in low density areas of the graph, we need to look for low-support trusses; Figure 9(c), for example, shows the strong 5-trusses. Conversely, if we wish to see the tightest cluster, which has support 15, we need to look for strong 17-trusses, as in Figure 9(e).

If, for every region of the graph, we instead show the strong truss of highest "local" support, we can produce the clusters shown in Figure 9(f); these are the strong "summit" trusses.

---

[8] https://snap.stanford.edu/data/ego-Facebook.html

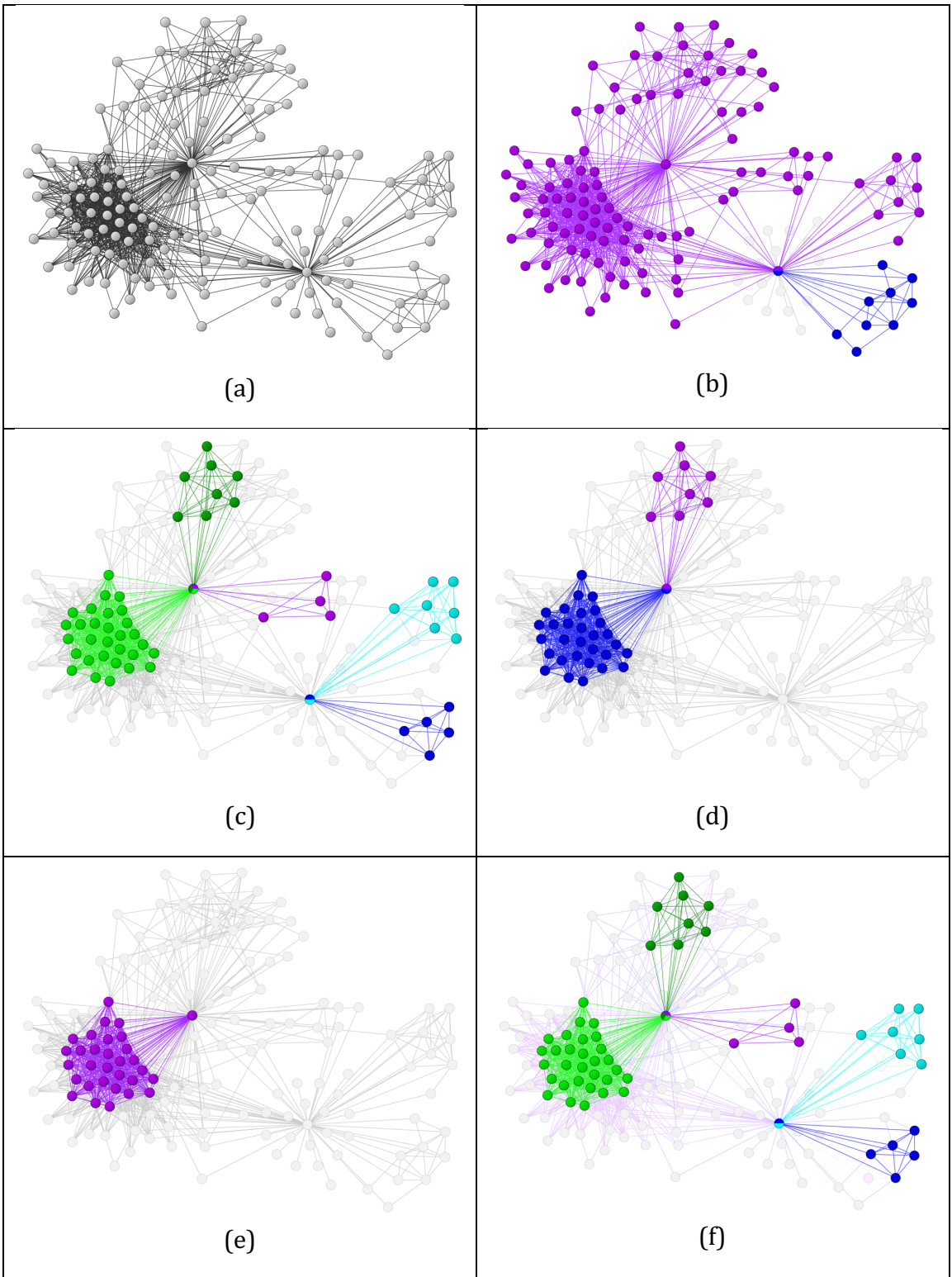

Figure 9. Subgraph of Facebook friendships drawn from McAuley and Leskovec (2012) showing strong trusses. (a) Graph, (b) 3-trusses, (c) 5-trusses, (d) 8-trusses, (e) 17-trusses, (f) summit trusses.

For motivation, we can consider a graph laid out in two horizontal dimensions, and imagine a surface above it whose height indicates tightness of vertex clusters. As suggested by the name, summit trusses occupy the local peaks in this terrain.

**Definition:** A summit truss is a maximal truss whose edges are not in a truss of higher support.

**Definition:** A strong summit truss is a maximal strong truss whose edges are not in a strong truss of higher support.

**Observation:** Summit trusses are edge-disjoint, as are summit strong trusses.

One can define summit trapezes in a parallel way.

Note that the summit trusses can be weighted or unweighted.

By computing summit trusses, we can discover locally tight groups throughout a graph, and without having to specify a support level. Moreover, showing only summit trusses permits a single picture to convey clusters in both dense and sparse areas of the graph, offering an overview without recourse to any parameters.

# Finding Trusses

One of the features that makes trusses practical is their ease of computation. This section will outline a method whose worst-case work scales with $m^{1.5}$, where $m$ is the number of edges in the graph.

Finding the trusses entails three steps: enumerating the triangles[9], trimming the triangle edges that lack sufficient support, and finding the components of the resulting graph.

Triangles may be found using one of the many variations of the method that Berry, et al. [37] refer to as "MinBucket," in which triangles are constructed from their minimum-degree vertex. One variation proceeds this way: each edge is examined in turn and recorded (in a "bucket") under the adjacent vertex whose degree is lowest (ties are decided in any consistent fashion). For each such bucket, each possible pair of edges in that bucket is considered to see if a closing edge exists to form a triangle. (We presume that the graph edges have been recorded such that each query can be done in O(1) time.[10])

Choosing to record each edge under its vertex of smallest degree minimizes the quadratic work that must be done in each bucket, making the worst-case running time[11] to be O($m^{3/2}$). Even better, Berry, et al. (2014) show that expected work for power law graphs of power at least 7/3 is linear in the number of edges.

Details and variations of this method can be found in Chiba and Nishizeki (1985), Schank and Wagner (2005), and Cohen (2008).

The result of enumerating triangles is that for each edge $e$, one now has the triangle support value sup($e$), the number of triangles containing that edge.

---

[9] Actual enumeration is not really necessary. All that is needed it to know the number of triangles containing each edge.
[10] A simple hash table with the vertex pair as a key suffices.
[11] To see this, partition the vertices into set $L$, having degrees less than or equal to $m^{1/2}$, and set $H$, containing the remaining high-degree vertices. The work to find the triangles goes as the sum of the squares of the bucket sizes $\{b_i\}$. For vertices in $L$, $\sum b_i^2 \leq m^{1/2} \sum b_i \leq m^{3/2}$. For vertices in $H$, $b_i < |H|$ so that $\sum b_i^2 < |H|^3$. But the number of vertices in $H$ is bounded by the sum of their degrees, which must not exceed twice the number of edges: $|H| < 2m/m^{1/2} = 2m^{1/2}$.

Wang and Cheng (2012) define the "trussness" $\phi(e)$ of edge $e$ as 2 plus the highest support of a truss containing $e$, and the corresponding "k-class" $\Phi_k = \{e \in E(G) | \phi(e) = k\}$. Moreover, they offer a nice algorithm for computing the trussness of all edges in $G$ whose worst-case computation is $O(m^{3/2})$. Their algorithm is shown as Algorithm 1.

---

**find_k_classes**(graph $G$):

    $k \leftarrow 2$;
    $\forall e \in E(G), s(e) \leftarrow \sup(e)$;
    sort edges in ascending order of $s$;
    while ( $E(G) \neq \emptyset$ )
    {
    $\Phi_k \leftarrow \emptyset$;
        while ( $\exists e \ni s(e) \leq k - 2$ )
        {
            let $e = (u, v)$ be edge with lowest $s(e)$;
            assume, w.l.o.g., deg $(u) \leq$ deg $(v)$;
            for ( $w \in$ neighbors$(u)$ )
            {
                if ( $(v, w) \in E(G)$ )
                {
                    $s\big((u, w)\big) \leftarrow s\big((u, w)\big) - 1$;
                    $s\big((v, w)\big) \leftarrow s\big((v, w)\big) - 1$;
                    reorder $(u, w)$ and $(v, w)$ by new $s(e)$;
                };
            };
            $\Phi_k \leftarrow \Phi_k \cup \{e\}$;
            $G \leftarrow G - e$;
        };
        $k \leftarrow k + 1$;
    };
    return $\{\Phi_k\}$;

Algorithm 1: Algorithm for Finding all k-classes in graph $G$ due to Wang and Cheng (2012).

---

Due to the degree ordering, the innermost loop will be executed $O(m^{1.5})$ times; the argument is similar to the one for triangle enumeration. The initial sorting is linear in the number of edges, being conducted, for example, by radix sort. Wang and Cheng suggest a structure for holding the edges sorted by $s()$ such that the subsequent reordering can be done in constant time. Overall, Algorithm 1 has the same complexity as enumerating the triangles. Note that in practice one does not really remove edges from $G$; they are merely marked as being removed.

Given the k-classes, one may determine the k-trusses for some $k$ by finding the components of the subgraph induced by $\bigcup_{i \geq k} \Phi_i$; component finding is linear in the number of edges involved. One may also construct the entire "dendrogram" of trusses with linear work, regarding them as agglomerative clusters formed by single-link agglomeration: Let $k_{\max} = \max\{k | \Phi_k \neq \emptyset\}$. Nontrivial clusters at the $k_{\max}$ level are simply the components of the graph induced by $\Phi_{\max}$. These, plus the implicit clusters consisting of single nodes, form the starting clusters (though one is only interested in clusters that are not isolates). One may then add the edges in $\Phi_{\max-1}$ one at a time; at completion, the resulting clusters (ignoring the isolates) are the $k_{\max-1}$-level clusters and $k_{\max-1}$-trusses. One can continue this process until the level $k = 2$ is completed. (See Finding the Strong Trusses below.) The combined work is linear in the number of edges.

# Finding Strong Trusses

The strong trusses may be found for individual levels from the corresponding (weak) trusses, or may be found all at once as a family of agglomerative clusters once the k-classes have been determined.

An example approach is outlined in Algorithm 2, where $C_e$ indicates the cluster containing edge $e$. Edges are added from the k-classes starting at the maximum, each being assigned an initial singleton cluster. For each edge, we ask if that edge forms a triangle with already-added edges; if so, clusters containing those edges are merged with the cluster containing the new edge. The level of these joins, equal to the k-class of the new edge, is noted. The set $D$ keeps track of which edges have appeared; this can be implemented merely by flagging edges in any way that permits constant time for testing membership.

```
find_strong-trusses(graph G):

    kD ← ∅;
    for ( k: k_max down to 2 )
    {
        for ( e = (u, v) ∈ Φ_k )
        {
            create cluster C_e ← {e};
            assume, w.l.o.g., deg (u) ≤ deg (v);
            for ( w ∈ neighbors(u) )
                if ( (u, v) ∈ D  and  (v, w) ∈ D )
                    merge C_e, C_(u,v), and C_(v,w) at level k;
            D ← D ∪ {e};
        };
    };
```

Algorithm 2: Sketch of finding all strong trusses as a family of agglomerative clusters

The function neighbors($u$) can be implemented on the original graph or on a new graph or adjacently list being accumulated while adding edges in Algorithm 2. As earlier, searching neighbors of the smaller degree vertex results in a total work of $O(m^{1.5})$ to construct the entire family of clusters.

Figure 10 shows the dendrogram of clusters resulting from Algorithm 2 applied to the graph in Figure 10. At level $k = 5$ (support 3), there is one strong truss of 10 edges. At level 4, a new strong truss of 5 edges is formed (and the strong truss found earlier remains separate). Level 3 sees these two trusses combined with 3 more edges to form a single strong truss.

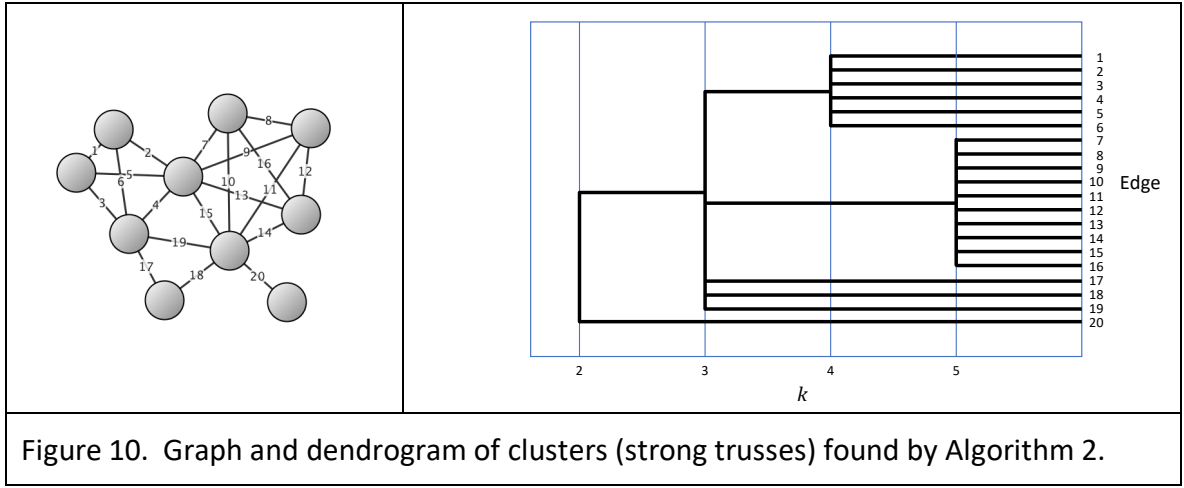

Figure 10. Graph and dendrogram of clusters (strong trusses) found by Algorithm 2.

# Finding Weighted Trusses

Finding the weighted trusses is much like finding the unweighted ones. We now use the weighted support values, which are still integers. As the process proceeds, an edge loses weighted support, meaning that its support value is decremented by the weight of a triangle when a supporting edge is removed; that weight may be larger than 1. Notionally, we can keep track of each edge's support by placing it in a set for that support value and keeping a map $M$ that maps the edge to its support value. We will have one set $Q_i$ for each possible support value $i$.

The modified algorithm is shown in Algorithm 3.

**find_weighted_k_classes**(graph $G$):

    $\forall e \in E(G)$, compute weighted support $S_w(e)$;
    $\forall i \in \{0, 1, \ldots, \max S_w(e)\}$, set $Q_i \leftarrow \emptyset$;
    $\forall e \in E(G)$
    {
        $M(e) \leftarrow S_w(e)$;
        $Q_{M(e)} \leftarrow Q_{M(e)} \cup \{e\}$;
    }
    $k \leftarrow 2$;
    while ( $E(G) \neq \emptyset$ )
    {
    $\Phi_k \leftarrow \emptyset$;
        while ( $Q_{k-2} \neq \emptyset$ )
        {
            pull $e = (u, v)$ from $Q_{k-2}$;
            assume, w.l.o.g., deg $(u) \leq$ deg $(v)$;
            for ( $w \in$ neighbors$(u)$ )
            {
                if ( $(v, w) \in E(G)$ )
                {
                    $\delta \leftarrow W_\Delta\big(e, (u, w), (v, w)\big)$;
                    remove $(u, w)$ from $Q_{M((u,w))}$;
                    remove $(v, w)$ from $Q_{M((v,w))}$;
                    $M\big((u, w)\big) \leftarrow \max\{M\big((u, w)\big) - \delta, k - 2\}$;
                    $M\big((v, w)\big) \leftarrow \max\{M\big((v, w)\big) - \delta, k - 2\}$;
                    $Q_{M((u,w))} \leftarrow Q_{M((u,w))} \cup \{e(u, w)\}$;
                    $Q_{M((v,w))} \leftarrow Q_{M((v,w))} \cup \{e(v, w)\}$;
                };
            };
            $\Phi_k \leftarrow \Phi_k \cup \{e\}$;
            $G \leftarrow G - e$;
        };
        $k \leftarrow k + 1$;
    };
    return $\{\Phi_k\}$;

Algorithm 3: Algorithm for Finding all weighted k-classes in graph $G$

The inner loop is executed $O(m^{1.5})$ times, as in previous algorithms. Work to remove an edge from a set or add one to set is constant time. Overall work, then, is $O(m^{1.5} + \max S_w(e))$; the latter term due to maximum weighted support is not likely to be significant.

Finding strong weighted trusses from the weighted k-classes is the same as finding strong trusses from unweighted k-classes.

# Finding Summit Trusses

Armed with the dendrograms described above, finding summit trusses is simple: one merely selects those clusters (trusses) that have no cluster ancestors other than trivial (single edge) clusters. The dendrogram of Figure 10 shows 2 summit trusses, one with $k = 4$ and one with $k = 5$.

# Finding Trapezes

The steps for finding trapezes are slightly more complicated, but related to those for finding trusses.

Let us order all of the vertices by degree, resolving ties consistently, say by using the name of the vertex; let $s(v)$ map vertex $v$ to its order.

For each edge $e = (u, v)$, with $s(u) < s(v)$, we bin $e$ in a low degree bin $L_u$ and a high degree bin $H_v$. At the end of this process, for each vertex $v$, $L_v$ contains all edges whose low degree vertex is $v$ and $H_v$ contains all edges whose high degree vertex is $v$. For $m$ edges, this takes $O(m)$ work.

Before proceeding, let us establish some terminology: Two edges $(a, u)$ and $(a, v)$ with a common vertex $a$ will be called an open triad with apex $a$ and peripheral vertices $u$ and $v$. For simplicity, we will omit the word "open," and merely use the term "triad." For $s(u) < s(v)$, we will say that this triad has periphery $(u, v)$. In the event that $s(a) < s(u)$ and $s(a) < s(v)$, we will refer to the triad as a low-apex triad; if $s(u) < s(a) < s(v)$ holds, then we will call it a median-apex triad. (The third case is one we shall avoid.)

Let us consider a rectangle constructed from vertices $v_1$, $v_2$, $v_3$, and $v_4$ with $s(v_1) < s(v_2) < s(v_3) < s(v_4)$. Figure 11 illustrates the three possibilities. In each case, the rectangle can be constructed by two triads, each of which is either a low-apex triad or a median-apex triad. Indeed, this representation is unique: exactly one such construction exists for each 4-cycle in the graph. We will say that a triad that is either a low-apex triad or a median-apex triad is admissible.

The act of counting rectangles, then, is accomplished by looking for common peripheries of admissible triads. If $n$ admissible triads have the same periphery, then we can construct $n(n-1)/2$ rectangles from them — one for each distinct pair.

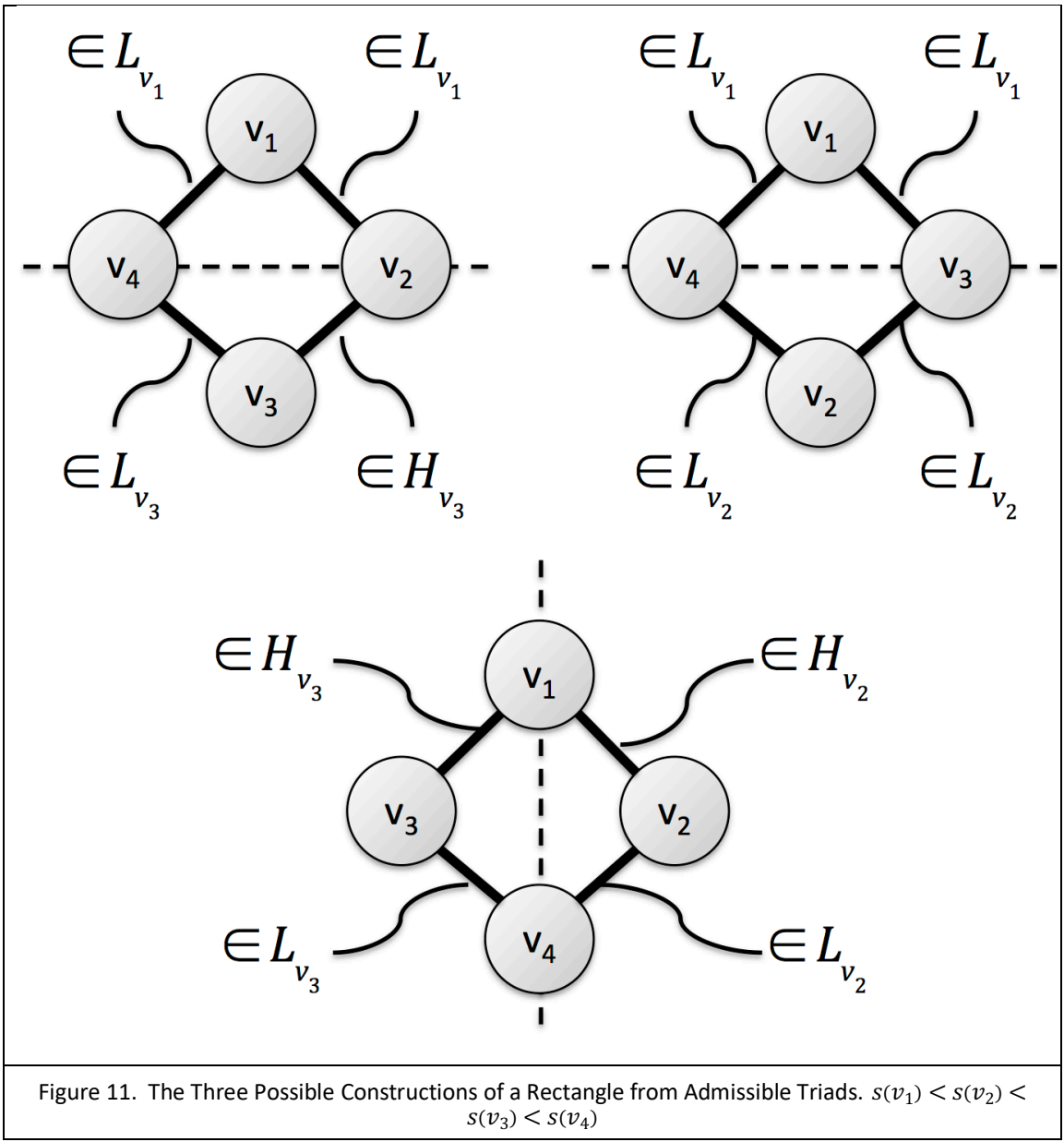

Figure 11. The Three Possible Constructions of a Rectangle from Admissible Triads. $s(v_1) < s(v_2) < s(v_3) < s(v_4)$

These observations lead to a prescription for finding trapezes in graph $G$. We can construct an edge-triad-periphery graph $T$ that contains three types of nodes: those that represent edges in $G$, those that represent admissible open triads in $G$ containing those edges, and those that represent the peripheries of those open triads. The graph $T$ will serve to count the number of rectangles containing each edge, and allow us, ultimately, to construct trapezes.

For each vertex $v$, we consider each pair of edges $(v, a)$ and $(v, b)$ in $L_v$ and record a map of both of those edges to a triad and then to the triad's periphery $(a, b)$ or $(b, a)$ (putting them in order), by adding an edge in $T$ from a node representing $(v, a)$ to a node $t$ representing the triad $(a, v, b)$, a similar edge from a node representing $(v, b)$ to $t$, and finally an edge from $t$ to a node representing the periphery $(a, b)$ or $(b, a)$. We also consider each pair of edges obtained by drawing one from $L_v$ and one from $H_v$ and making similar constructions in $T$.

When we are done with this operation, we will have an edge-triad-periphery graph $T$, in which the periphery of each admissible triad of edges in $G$ is connected to vertices representing triads having that periphery; the triad vertices will also have edges to vertices in $T$ representing edges of $G$ that constitute the triads. A periphery vertex in $T$ having degree of $d$ will connect to $d$ open triads and will represent $n(n-1)/2$ rectangles. Any periphery vertex with degree 1 represents no rectangles, and will be removed. Edges in $T$ can be regarded as directional, leading from the edge vertices to the triad vertices, thence to the periphery vertices.

For any edge $e$ in $G$, we can obtain the number of rectangles involving $e$ by following the edges forward in $T$ to a set of descendant periphery vertices $\{p_i\}$ in $T$. The number of rectangles containing $e$ is $\sum(d(p_i) - 1)$, where $d(v)$ denotes the degree of vertex $v$.

The work to construct this graph, proportional to the number of open triads constructed, is bounded by $O(m^{3/2})$, where $m$ is the number of edges in $G$: Let the set $S = \{v | d(v) \leq \sqrt{m}\}$ contain the vertices of small degree, with the set $B$ containing the remaining vertices of big degree. For each vertex $v$, let $l_v = |L_v|$ and $h_v = |H_v|$. The work to record the open triads whose apexes are in $S$ is proportional to

$$\sum_{v \in S}(l_v^2 + l_v h_v) = \sum_{v \in S} l_v(l_v + h_v) = \sum_{v \in S} l_v d_v \leq \sqrt{m} \sum_{v \in S} l_v \leq m^{3/2}.$$

Since the degrees of vertices in $B$ are greater than $\sqrt{m}$, and the sum of degrees cannot exceed $2m$, $|B| < 2\sqrt{m}$. The work to record the open triads whose apexes are in $B$ is proportional to

$$\sum_{v \in B} l_v d_v < \sum_{v \in B} |B| d_v = |B| \sum_{v \in B} d_v < 4m^{3/2}.$$

The point of building a graph is not merely to count the number of rectangles containing each edge; that can be done more simply (but with the same complexity) merely by mapping each edge to the peripheries of admissible triads containing it to counters in a hash table. We build $T$ (or its equivalent) to find the trapezes.

Figure 12 offers a graph whose edges and vertices are labeled; Figure 13 is the corresponding edge-triad-periphery graph. (In constructing Figure 13, when two vertices had the same degree, the "lower" one was chosen as the one whose label was lexicographically first.)

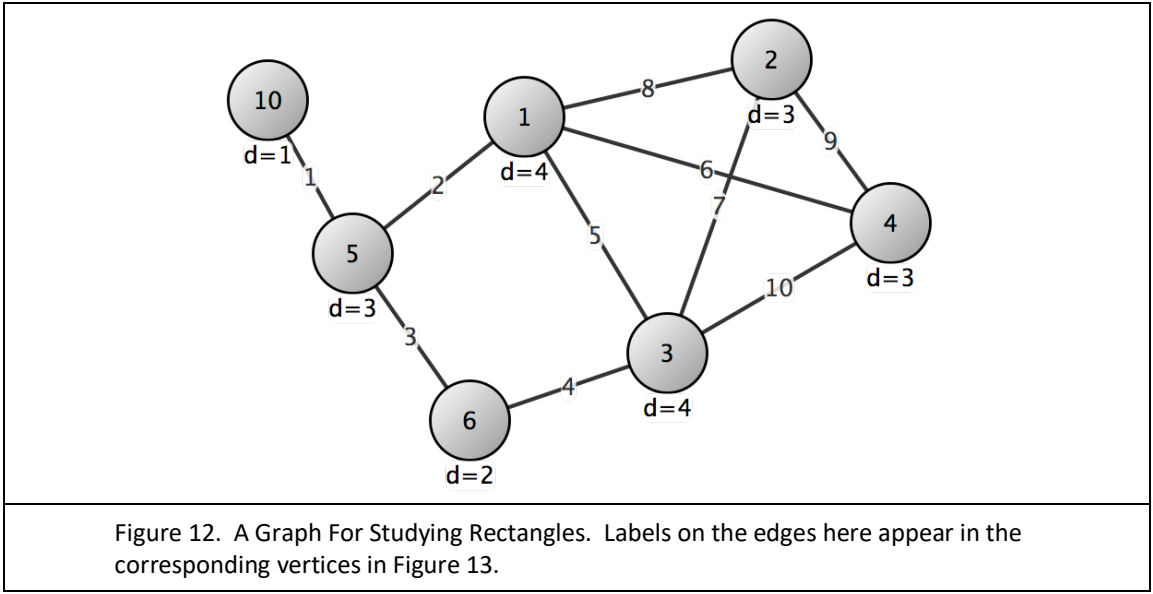

Figure 12. A Graph For Studying Rectangles. Labels on the edges here appear in the corresponding vertices in Figure 13.

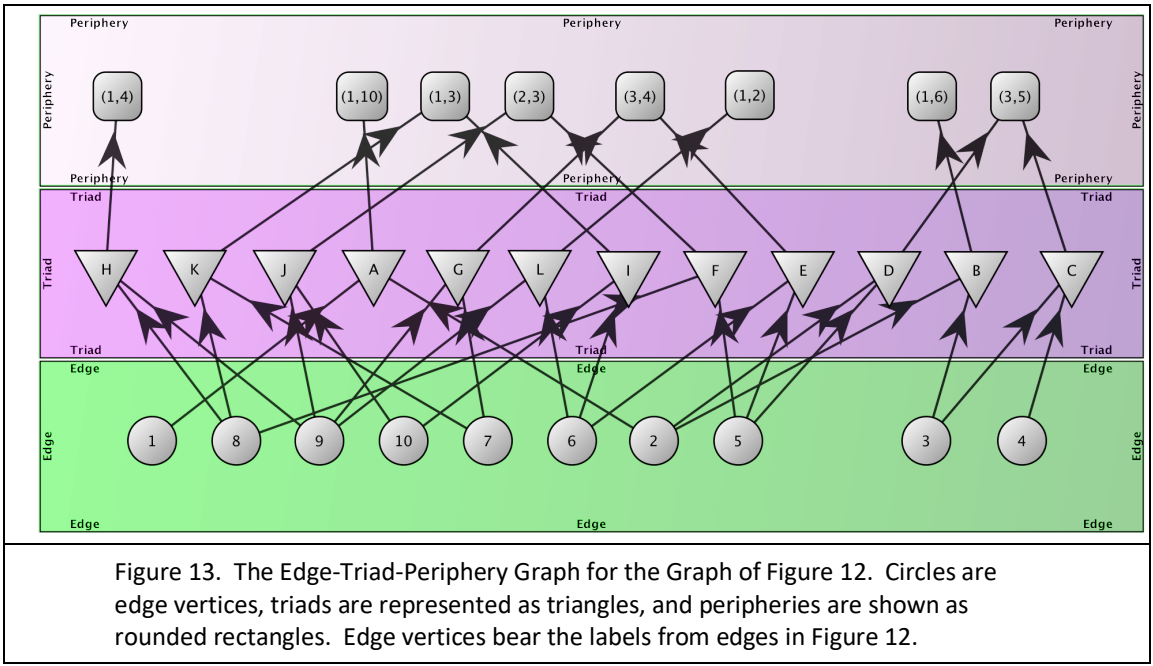

Figure 13. The Edge-Triad-Periphery Graph for the Graph of Figure 12. Circles are edge vertices, triads are represented as triangles, and peripheries are shown as rounded rectangles. Edge vertices bear the labels from edges in Figure 12.

Periphery vertices with a degree of 1 will not lead to rectangles; we can remove them and the triad vertices that lead to them. If that orphans any edge vertices, those can be removed, as well. Figure 14 shows the result of such trimming of the edge-triad-periphery graph of Figure 13.

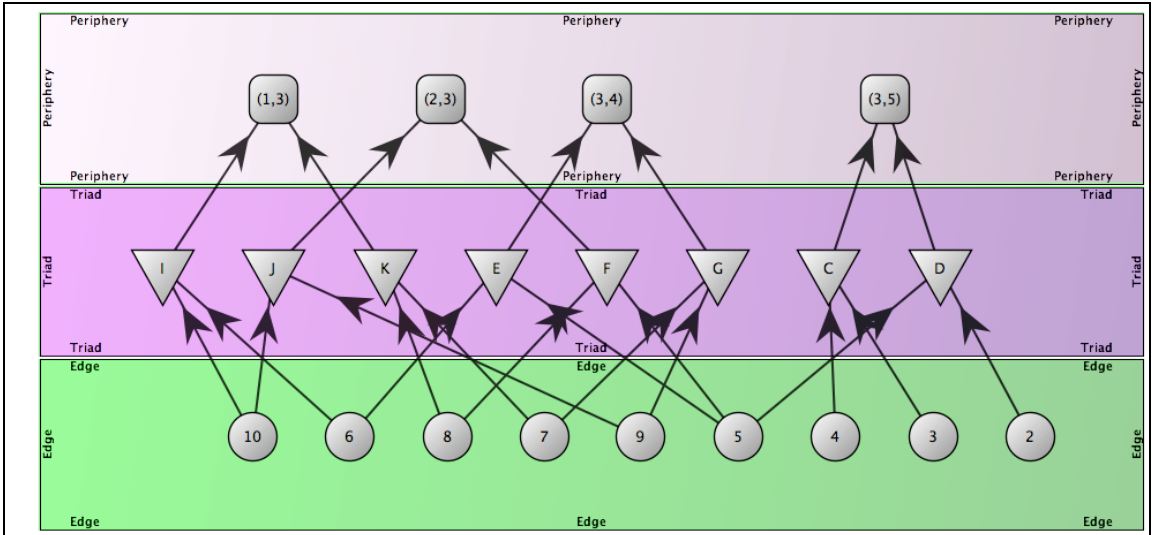

Figure 14. The Edge-Triad-Periphery Graph for the Graph of Figure 12 with Unproductive Vertices Removed. Circles are edge vertices, triads are represented as triangles, and peripheries are shown as rounded rectangles.

From Figure 14, we see that there are four rectangles, each made with a different periphery. We also see that edges 6, 7, 8, and 9 are in two of them and edge 5 is in three. In general, to record the counts of the number of rectangles each edge participates in, we propagate the contributions from each periphery vertex through the triad vertices, and sum them at each edge vertex; the contribution from a periphery vertex of degree $d$ is $d - 1$.

Having built the edge-triad-periphery graph and trimmed the nodes that obviously do not contribute, we can use it to determine the trapezes. Before doing so, we will make several definitions:

For any edge vertex $e$ in $T$, let $P(e)$ be the set of periphery vertices in $T$ that are descendants of $e$. In Figure 14, $P(10) = \{(1,3), (2,3)\}$. Conversely, for any periphery vertex $p$, let $E(p)$ be the set of edge vertex ancestors of $p$ in $T$. In Figure 17, $E((3,5)) = \{2, 3, 4, 5\}$.

We define for each edge vertex $e$ the tentative support $S(e)$ and for each periphery vertex $p$ the last reported support $Q(p)$. $Q(p)$ is initialized to $d(p) - 1$, and represents the last value of support that $p$ has reported to edges that are its ancestors in the graph $T$. $S(e)$ is our current understanding of the support of edge $e$, and is initialized to $S(e) = \sum_{p \in P(e)} Q(p)$.

We are now ready to find the trapezes of support $k$. We will do so by trimming $T$ until only those edge vertices that have support $k$ remain. We can then find the components of the subgraph induced by the edges represented by those remaining edge vertices. Trimming proceeds by alternating two phases: removing edge vertices that obviously lack sufficient support, and examining peripheries that had been driven by those removed vertices to update values of $Q$ and $S$ and see if more edges should be marked for removal.

The procedure for trimming $T$ is shown at a high level in Algorithm 4(a). The process begins by establishing the values of $Q$ and $S$. For each edge vertex $e$, the value of $S(e)$ is either the exact support of $e$ based on the currently-present set of edge vertices, or is an overestimate that will not alter the outcome of comparing it to $k$; the initial values established by step T3 are exact. A set $V$ of edge vertices to be deleted is created and populated by filling it in in step T5 with each edge vertex whose support is insufficient.

---

**trim**( edge-triad-periphery graph $T$, support $k$ )

(T1)   if not yet initialized:
       {
(T2)       for each periphery vertex $p$, set $Q(p) \leftarrow d(p) - 1$;
(T3)       for each edge vertex $e$, set $S(e) \leftarrow \sum_{p \in P(e)} Q(p)$;
       };

(T4)   let $V \leftarrow \emptyset$ be a set that contains edge vertices that must be removed;
(T5)   for each edge vertex $e$, if $S(e) < k$, then $V \leftarrow V \cup \{e\}$;

(T6)   while ( $V \neq \emptyset$ )
       {
(T7)       let $U \leftarrow \emptyset$ be a set that contains peripheries to be examined;
(T8)       for each $e \in V$, **removeEdgeVertex**( $e$ ) and remove $e$ from $V$;
(T9)       for each $p \in U$, **examinePeriphery**( $p$ );
       };

Algorithm 4(a): Main Routine For Trimming $T$. This may be called with increasing values of $k$; the steps T2 and T3 are only performed the first time.

---

With initialization out of the way, the procedure's main loop, steps T6 through T9, runs until there are no more edge vertices to delete, that is, when $V$ becomes empty. The current contents of $V$ are serviced in step T8, which removes each vertex in $V$ from the graph $T$, causing affected periphery vertices to be put into set $U$ for later examination, and removing adjacent triad vertices, which can cause updates to values of $S$ for the edge vertices that share those triads, or removal of those other edge vertices, if the removed triads are the last ones involving edge vertices. The details of the procedure **removeEdgeVertex** are given in Algorithm 4(b).

After set $V$ has been emptied in step T8, the periphery vertices affected by removal of edge vertices are dealt with in turn in step T9. Here, values of $Q$ and $S$ are updated, new vertex edges lacking support are placed in $V$ for removal, and obviated periphery and triad vertices are removed from $T$. The details are in **examinePeriphery**, described in Algorithm 4(c).

---

**removeEdgeVertex**( edge vertex $e$ ):

(R1)  for ( each triad $t$ of the form $t = (e, e')$ or $t = (e', e)$ )
      {
(R2)       add the periphery vertex child $p$ of $t$ to set $U$;

(R3)       if $d(e') = 1$, remove $e'$ from the graph $T$, else:
           {
(R4)            $S(e') \leftarrow S(e') - Q(p)$;
(R5)            If $S(e') < k, V \leftarrow V \cup \{e'\}$;
           };
(R6)       remove $t$ from $T$;
      };

(R7)  remove $e$ from $T$;

Algorithm 4(b): Routine for Removing An Edge Vertex from $T$.

---

The procedure for removing an edge vertex $e$ is outlined in Algorithm 4(b). Before removing $e$ from $T$, the triad vertices, children of $e$ in $T$, are dealt with in turn. For each such triad vertex $t$, its child vertex (a periphery) is marked for later examination in step R2, and it's other parent edge vertex $e'$ must be examined. If $t$ was the only triad involving $e'$, then $e'$ will no longer be involved, and it can simply be removed from $T$ (step R3); otherwise, the value of $S(e')$ will need to be updated to remove the support that had been offered to $e'$ through triad $t$. If the updated support is insufficient, $e'$ will be slated for removal by placing it in $V$. After these considerations, the triad vertex $t$ is removed. Finally, the edge vertex $e$ is removed from $T$.

The procedure for dealing with periphery vertex $p$ whose parent edge vertices have changed is outlined in Algorithm 4(c). If $p$ has been orphaned, then it is merely removed. Otherwise, we must update the support values of edge vertices that are ancestors of $p$ and remove $p$ (and its parent triad vertex) if it has only one parent triad vertex, since it can make no 4-cycles in that case.

If the degree of $p$ is large, the act of updating support values of ancestor edge vertices could be a big time consumer. But we make this observation: if the support $s$ that this periphery vertex offers its contributing edge vertices $E(p)$ is at least $k$, then we know that $S(e) \geq k\ \forall e \in E(p)$, and no edges in $S(e)$ will be removed, so we will not be asked to examine $p$. *Deus ex machina*: just when the large degree would cause expensive updating, the updating is not needed.

To update our contribution to each $S(e), \forall e \in E(p)$, we increment them by $\Delta = s - Q(p)$, that is, the last value that was propagated, $Q(p)$, is subtracted and the new value $s$ is added. Then the value of $Q(p)$ is updated to $s$ to indicate that it was the last value sent.

---

**examinePeriphery**( periphery vertex $p$ ):

(E1)   if $d(p) = 0$, remove $p$ from $T$ and leave;
(E2)   support of $p$: $s \leftarrow d(p) - 1$;
(E3)   let $\Delta \leftarrow s - Q(p)$;
(E4)   $Q(p) \leftarrow s$;

(E5)   for $e \in E(p)$:
       {
(E6)           $S(e) \leftarrow S(e) + \Delta$;
(E7)           if $S(e) < k, V \leftarrow V\{e\}$;
       };

(E8)   if $s = 0$:
       {
(E9)           remove parent triad $t$ of $p$ from $T$;
(E10)          remove $p$ from $T$;
       };

Algorithm 4(c): Routine for Examining a Periphery Vertex in $T$.

---

In practice, we may want to determine the trapezes for many support values $\{k_1, k_2, \ldots, k_{\max}\}$, where $k_1 < k_2 < \cdots < k_{\max}$. We can do this by calling **trim** with support $k_1$, using the surviving edges in $T$ to find the trapezes or strong trapezes, then repeating the process for successively larger values of support. Note that the initialization steps M2 through M3 are done only for the first call to **trim**.

It is likely that we will not want to examine every possible value of support; consider only support values of $2^i, i = 0, 1, \ldots i_{max}$.

We turn now to the matter of complexity. The number of periphery vertices that each edge can cause to be examined is limited: from the discussion of constructing triads, we see that the number of triads involving any edge must be less than $2\sqrt{m}$. Since each edge vertex can be removed once, at most, over *all* values of $k$ examined, **examinePeriphery** can be called at most $2m^{3/2}$ times. As remarked above, **examinePeriphery** can update at most $k$ vertex nodes, and the work to execute **examinePeriphery** one time is, at worst, $O(k)$. So over all calls to **trim**, the cost of step M9 is, at worst, $O(k_{max}m^{3/2})$. Similarly, the work to perform **removeEdgeVertex** one time is, at worst, $O(\sqrt{m})$, so the work to perform step M8 over all calls to **trim** is $O(m^{3/2})$, at worst. The initialization steps M2 and M3 take work proportional to the number of peripheral vertices, which scales, at worst, as $O(m^{3/2})$. We conclude that the worst case work for trimming is $O(k_{max}m^{3/2})$.

The work to build the trapezes from $T$ after one call to **trim** is $O(m)$. Since we will do this a maximum $k_{max}$ times, the work to execute **trim** dominates.

The space requirements for holding the triad vertices in $T$ dominate memory usage, and scale, at worst, as $O(m^{3/2})$.

# A Planted Partition Truss Example

Here we consider a widely used — and highly contrived — example: the planted *l*-partition model (Condon and Karp, 2001), in which the graph consists of *l* groups of *n/l* nodes, and whose edges are stochastic. Each pair of vertices of the same group is joined by an edge with probability *p*, while each pair of nodes drawn from distinct groups is joined by an edge with probability *r*, with *p* > *r*. The aim, of course, is to recover the groups as clusters.

As is customary, rather than specify *r* directly, one may specify the mixing parameter *μ*, the expected fraction of edges that lead from any node to nodes of another group. A mixing parameter of zero means that all edges are internal to groups; a mixing parameter of 1/2 indicates that, for any given node, the expected number of adjacent intragroup edges is the same as the expected number of adjacent intergroup edges.

Recovery of the groups is judged by the normalized mutual information (NMI) between the planted cluster assignments and the grouping into *k*-trusses (Danon, *et al.*, 2005). The NMI compares the node cluster labels in the planted model with the node labels assigned by the clustering algorithm under test. When one perfectly predicts the other, that is, when there is a map of one label to the other, then the NMI is 1. When one does nothing to predict the other, then the NMI is 0. For purposes of computing the NMI, each actor not placed in a truss is assigned to its own cluster.

Table 1 shows results for treating recovered (weak) trusses as clusters. As might be expected, values of *k* that are too small fail to have sufficient resolution, that is, they are too indiscriminate, joining clusters that should be separate. Values of *k* that are too high expect too much and fail to place some vertices in clusters. The reported time[12] (in seconds) includes the time to record the results for all values of *k* present. The entries having a value of zero result from the entire graph being a single truss.

| MODEL | | | | AVE GRAPH | | | AVE RESULT OF CLUSTERING | | | | | | | | | | | |
|---|---|---|---|---|---|---|---|---|---|---|---|---|---|---|---|---|---|---|
| | | | | | | | Normalized Mutual Information | | | | | | | | | | | |
| *l* | *n/l* | *p* | *μ* | *n* | *m* | TIME (S) | *k* = 3 | 4 | 5 | 6 | 7 | 8 | 9 | 10 | 11 | 12 | 13 | 14 |
| 10 | 10 | 0.8 | 0.1 | 100 | 501 | .004 | 0.63 | 0.96 | 0.95 | 0.82 | 0.69 | – | – | – | – | – | – | – |
| 20 | 20 | 0.8 | 0.1 | 400 | 3.7k | .007 | 0 | 0.96 | 1.00 | 1.00 | 1.00 | 1.00 | 0.97 | 0.87 | 0.75 | 0.69 | 0.14 | – |
| 20 | 20 | 0.8 | 0.5 | 400 | 8.5k | .2 | 0 | 0 | 0 | .36 | 1.00 | 1.00 | 0.98 | 0.87 | 0.76 | 0.41 | 0.27 | – |
| 1k | 20 | 0.8 | 0.5 | 20k | 426k | 10 | 0 | 1.00 | 1.00 | 1.00 | 1.00 | 1.00 | 0.99 | 0.95 | 0.87 | 0.83 | 0.82 | 0.33 |
| 10k | 20 | 0.8 | 0.5 | 200k | 4.26M | 100 | 0.19 | 1.00 | 1.00 | 1.00 | 1.00 | 1.00 | 0.99 | 0.96 | 0.90 | 0.87 | 0.86 | 0.86 |

Table 1. Clustering of Planted *l*-Partition Graphs using (Weak) Trusses. All existing truss sizes are shown. Entries are averages over multiple trials. The number of vertices is *n*, the number of edges is *m*.

---

[12] This was done on a 15-inch MacBook Pro, issued in late 2013.

Table 2 is similar, but reports the results for strong trusses, which should do a better job when trusses are composed of relatively well-connected sections joined by a few nodes. Indeed, strong $k$-trusses outperform week $k$-trusses for small values of $k$, and have nearly identical performance for intermediate values; the comparison for larger values of $k$ is mixed.

| MODEL | | | | AVE GRAPH | | AVE RESULT OF CLUSTERING | | | | | | | | | | | |
|---|---|---|---|---|---|---|---|---|---|---|---|---|---|---|---|---|---|
| | | | | | | TIME (S) | Normalized Mutual Information | | | | | | | | | | |
| $l$ | n/l | $p$ | $\mu$ | $n$ | $m$ | | $k = 3$ | 4 | 5 | 6 | 7 | 8 | 9 | 10 | 11 | 12 | 13 | 14 |
| 10 | 10 | 0.8 | 0.1 | 100 | 426 | .006 | 0.93 | 1.0 | 0.96 | 0.86 | 0.76 | 0.42 | – | – | – | – | – | – |
| 20 | 20 | 0.8 | 0.1 | 400 | 3.6k | .07 | 0.81 | 1.0 | 1.00 | 1.00 | 1.00 | 0.99 | 0.98 | 0.90 | 0.75 | 0.68 | – | – |
| 20 | 20 | 0.8 | 0.5 | 400 | 8.5k | .5 | 0 | 0 | 0 | .97 | 1.00 | 1.00 | 0.97 | 0.87 | 0.75 | 0.41 | – | – |
| 1k | 20 | 0.8 | 0.5 | 20k | 426k | 10 | 0.86 | 1.00 | 1.00 | 1.00 | 1.00 | 1.00 | 0.99 | 0.95 | 0.87 | 0.83 | 0.82 | 0.66 |
| 10k | 20 | 0.8 | 0.5 | 200k | 4.26M | 100 | 0.99 | 1.00 | 1.00 | 1.00 | 1.00 | 1.00 | 0.99 | 0.96 | 0.90 | 0.87 | 0.86 | 0.86 |

Table 2. Clustering of Planted $l$-Partition Graphs using Strong Trusses. All existing truss sizes are shown. Entries are averages over multiple trials. The number of vertices is $n$, the number of edges is $m$.

Tables 3 and 4 are analogs of Tables 1 and 2, and report on recovering clusters via summit trusses and strong summit trusses, respectively. As can be seen from the tables, the summit trusses recover much of the cluster structure without recourse to specifying a support level. This comes at the price of not doing as well as the optimum level shown in Tables 1 and 2.

| MODEL | | | | AVE GRAPH | | AVE RESULT OF CLUSTERING | |
|---|---|---|---|---|---|---|---|
| $l$ | n/l | $p$ | $\mu$ | $n$ | $m$ | TIME (S) | Normalized Mutual Information |
| 10 | 10 | 0.8 | 0.1 | 100 | 420 | 0.005 | 0.91 |
| 20 | 20 | 0.8 | 0.1 | 400 | 3.6k | 0.07 | 0.94 |
| 20 | 20 | 0.8 | 0.5 | 400 | 8.5k | 0.2 | 0.93 |
| 1k | 20 | 0.8 | 0.5 | 20k | 430k | 8 | 0.97 |
| 10k | 20 | 0.8 | 0.5 | 200k | 4.3M | 80 | 0.98 |

Table 3. Clustering of Planted $l$-Partition Graphs using Summit Trusses. All existing truss sizes are shown. Entries are averages over multiple trials. The number of vertices is $n$, the number of edges is $m$.

| MODEL | | | | AVE GRAPH | | AVE RESULT OF CLUSTERING | |
|---|---|---|---|---|---|---|---|
| $l$ | $n/l$ | $p$ | $\mu$ | $n$ | $m$ | TIME (S) | Normalized Mutual Information |
| 10 | 10 | 0.8 | 0.1 | 100 | 420 | 0.01 | 0.90 |
| 20 | 20 | 0.8 | 0.1 | 400 | 3.6k | 0.06 | 0.92 |
| 20 | 20 | 0.8 | 0.5 | 400 | 8.5k | 0.4 | 0.93 |
| 1k | 20 | 0.8 | 0.5 | 20k | 430k | 10 | 0.93 |
| 10k | 20 | 0.8 | 0.5 | 200k | 4.3M | 100 | 0.97 |

Table 4. Clustering of Planted *l*-Partition Graphs using Summit Strong Trusses. All existing truss sizes are shown. Entries are averages over multiple trials. The number of vertices is *n*, the number of edges is *m*.

In cases of clusters that vary in size within the same graph, summit trusses often do as well, if not better, than fixed support detection of clusters. An example is offered in Table 5, in which planted clusters vary uniformly over sizes 5 through 10.

| Average of Normalized Mutual Information | | | | | | |
|---|---|---|---|---|---|---|
| $k$ = 3 | 4 | 5 | 6 | 7 | 8 | Summit |
| 0 | 0.50 | 0.85 | 0.77 | 0.72 | 0.70 | 0.91 |

Table 5. Clustering of Planted *l*-Partition Graphs using Trusses of fixed support and using Summit Trusses. Ten planted clusters with sizes ranging uniformly over {5, 6, …, 10}. Entries are averages over multiple trials.

# Some Real Truss Examples

The author constructed a graph from the friendship relations on Facebook of his 430 "friends" on Facebook. This example has the merit of being constructed of real relations that require reciprocation, though the ties lack any strength information, the relationships may be tenuous, and only going one "hop" from each of the author's friends restricts the analysis quite a bit. The most important feature of such a graph was that the author would have some idea of ground truth.

The result of this collection was a graph of 198k actors (171k of which were pendant) and 326k edges. The author removed himself from the graph.

The high number of actors results from the author being an amateur musician and composer and having many high-degree music friends. The expectation was that many of the musicians would form a tight group or multiple tight groups.

Figure 15 shows the portion of the graph that found itself in strong 4-trusses. Each strong truss has been collapsed into a single composite node, with the exception that actors who are in more than one strong truss are shown separately. In the figure, distinct colors denote distinct trusses and nodes are sized and labeled to indicate the number of actors each represents.

Not surprisingly, the bulk of the 9,391 actors represented here appear in a single strong truss for *Musicians*. The author's spouse is a connecting member of four strong trusses.

Three social circles appear. They represent geographically-separated groups with little cognizance of each other.

The two work circles represent groups drawn from distinct professions that do not much intermingle. But *Social Circle 2* derives from former professional colleagues, explaining the connection to *Work Circle 2* and the common member.

The two groups deriving from high school acquaintances, while not geographically distinct, have little interaction, save a single member who is in both.

Finally, those members of the author's family who use Facebook appear in a strong truss of their own.

Note that not only about 35% of the 27k non-pendant actors are represented in 4-trusses.

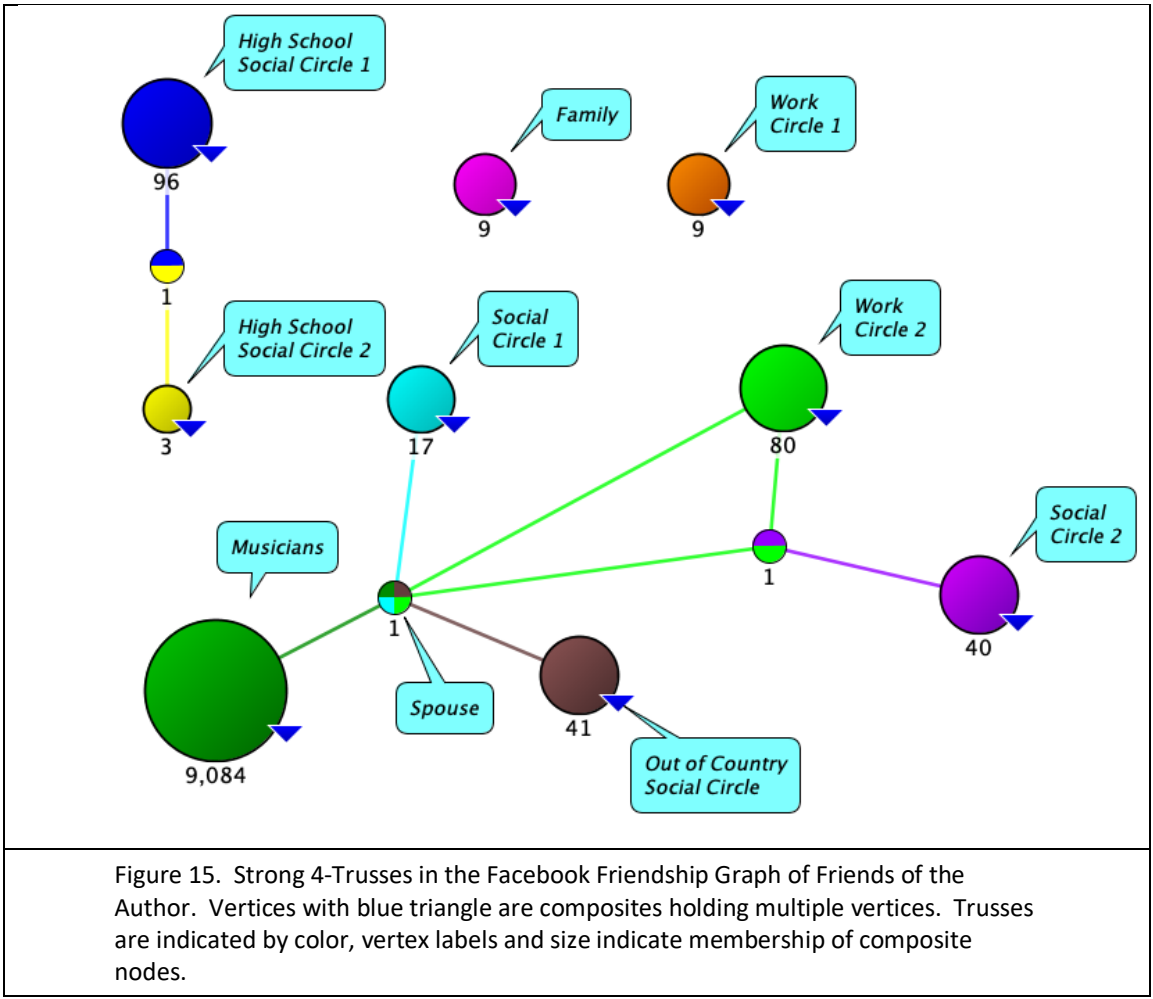

Figure 15. Strong 4-Trusses in the Facebook Friendship Graph of Friends of the Author. Vertices with blue triangle are composites holding multiple vertices. Trusses are indicated by color, vertex labels and size indicate membership of composite nodes.

Figure 16 shows a similar graph, but for all strong summit trusses of support greater than or equal to 2. These groups are necessarily wholly contained in those of Figure 15. While some strong 4-trusses were themselves summits (Family, Work Circle 1, Social Circle 2), most were not. There were four summits within the Musicians truss, accounting for only 80 of the 9,084 members of the original; all others exhibited single summits that were considerably smaller than their membership in Figure 15.

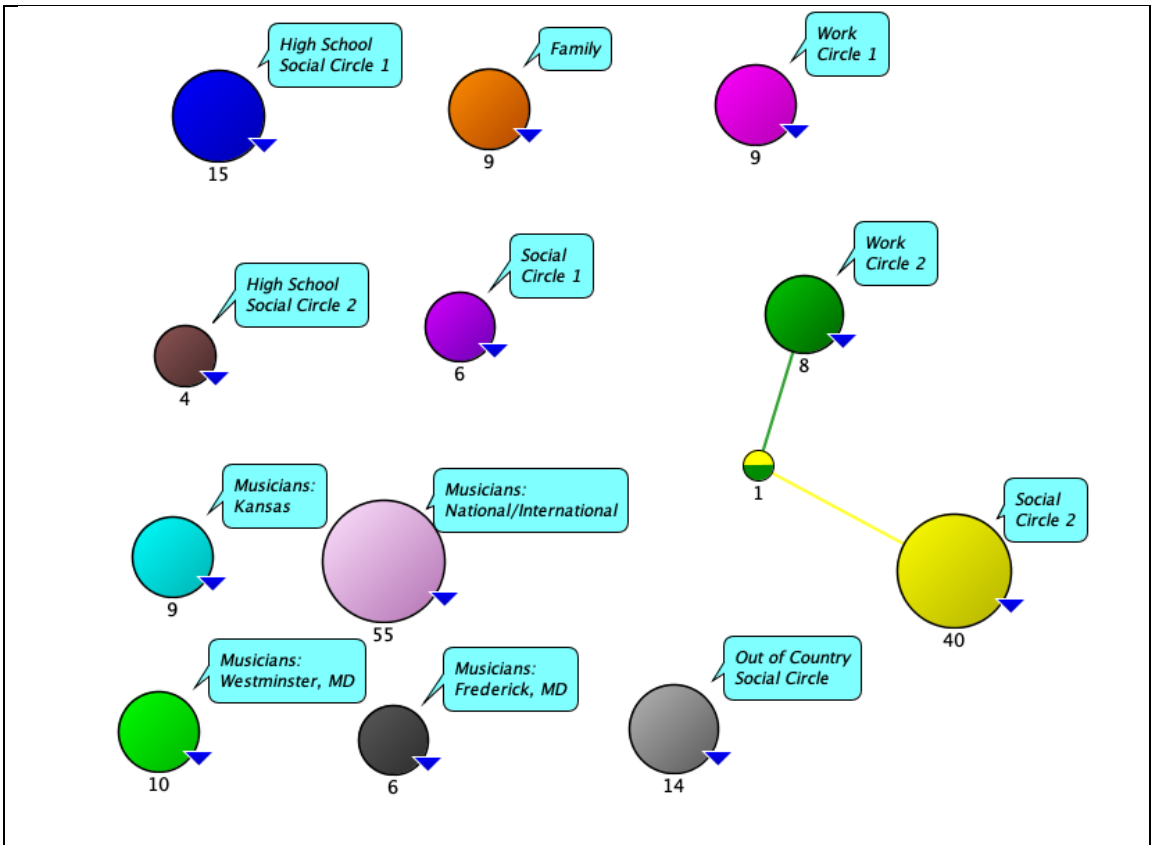

Figure 16. Strong Summit Trusses Facebook Friendship Graph of Friends of the Author, Support Greater than 1. Vertices with blue triangle are composites holding multiple vertices. Trusses are indicated by color, vertex labels and size indicate membership of composite nodes. Compare to Figure 15.

The results of this experiment agree with intuition: the trusses — and particularly strong trusses — identified professional and social communities; higher levels of truss support produced tighter subgroups within the more inclusive communities seen at lower support levels. In particular, geographically-distinct factions emerged from the larger musician strong truss at higher values of $k$. Examination of these actors verified that they were, indeed, sensible groupings.

Here is another example drawn from social media, this time using the platform Reddit. The question to be addressed was this: among those actors posting to r/healthcare, what other forums were they posting to, and what were the significant groups of such forums? The authors of posts (and comments) to the subreddit r/healthcare were determined and their posts to other subreddits were noted. A graph was then built consisting of vertices that represented subreddits receiving posts. An edge was constructed between two vertices if they shared at least one author; that edge recorded the number $c$ of common post authors. After dropping the r/healthcare vertex and the edges representing only one common author, the graph consisted of 275 vertices with 1340 edges. Much of the graph was likely insignificant: authors posting to nonspecific forums such as r/funny, r/pics, and r/news. An edge weight was constructed to reward edges with strong common author count and to penalize edges for adjacency to nodes representing subreddits with high subscribership $s$:

$$w((u,v)) = \tau\Big(c\big(e(u,v)\big)\Big) \min\{\omega(v), \omega(u)\},$$

where

$$\tau(c) = \begin{cases} 1, & c \leq 2 \\ 10, & c = 3 \\ 20, & c = 4 \\ 30, & c \geq 5 \end{cases}$$

and

$$\omega(v) = 1 + \ln\left(\frac{s_{\max}}{s(v)}\right)$$

and where $s_{\max} = \max\{s(v)\}$ over all vertices.

The result is pictured in Figure 17, which shows the summit trusses with support greater than 10. Since much of healthcare involves funding, there is a significant truss involving cryptofinance. There is a small truss restricted to healthcare professionals, some of whom are also answering questions raised in a truss of forums seeking advice. Note that the nonspecific forums (r/AskReddit, r/pics, etc.) are confined to a single truss.

This result would likely not have been achievable with weighting.

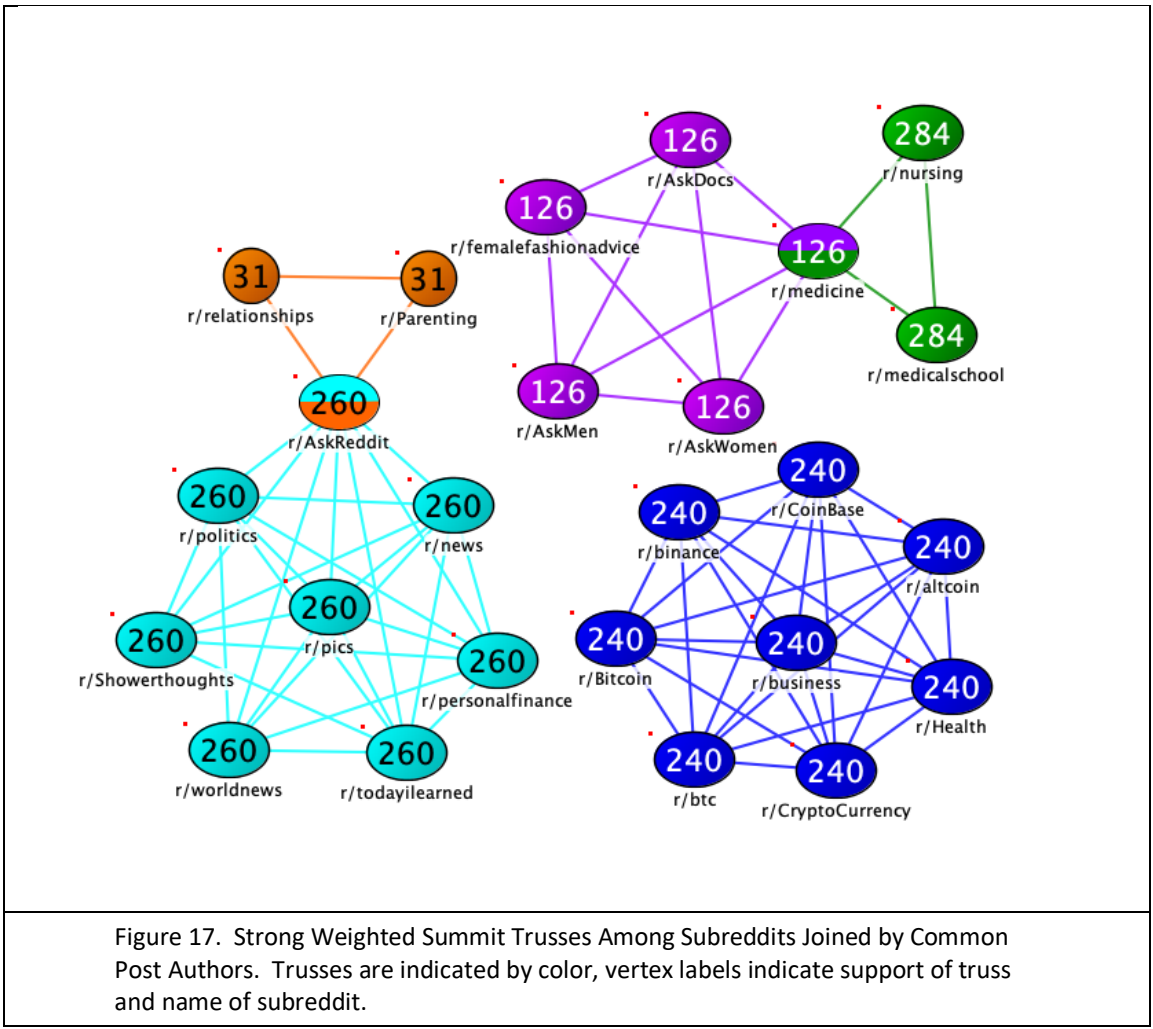

Figure 17. Strong Weighted Summit Trusses Among Subreddits Joined by Common Post Authors. Trusses are indicated by color, vertex labels indicate support of truss and name of subreddit.

# A Larger Real Trapeze Example

Brightkite was once a location-based social networking service provider where users shared their locations via check-ins. Some of the (anonymized) data has since been made available (Cho, *et al.*, 2011). This data consists of two sets: records that connect users with locations (check-ins), and an independent friendship graph resulting from user input. The check-ins graph contains 1.1M edges, 0.82M nodes (51k actors, 0.77M locations), and 4.7M check-ins over the period of March 2008 through October 2010.  It would seem that we could detect actors who are likely to know each other by looking for visits to the same rarely visited locations.  Accordingly, all locations that were visited by less than two actors or more than 10 were removed, then all edges representing less than 2 visits were also removed.  The remaining graph contained 89k nodes (21k actors and 68k locations) and 116k edges.  (Note that there is at most one edge between an actor and a location; it bears at attribute equal to the number of visits that it represents.)

Figure 18 shows an example of a 16-trapeze in the remaining check-in graph.  (There were 41 such 16-trapezes.)  This one involves three actors and 34 locations.  The number of common rarely-visited locations strongly suggests relationships between the actors here.  Indeed, the separate friendship graph shows that these three actors identified each other as friends.

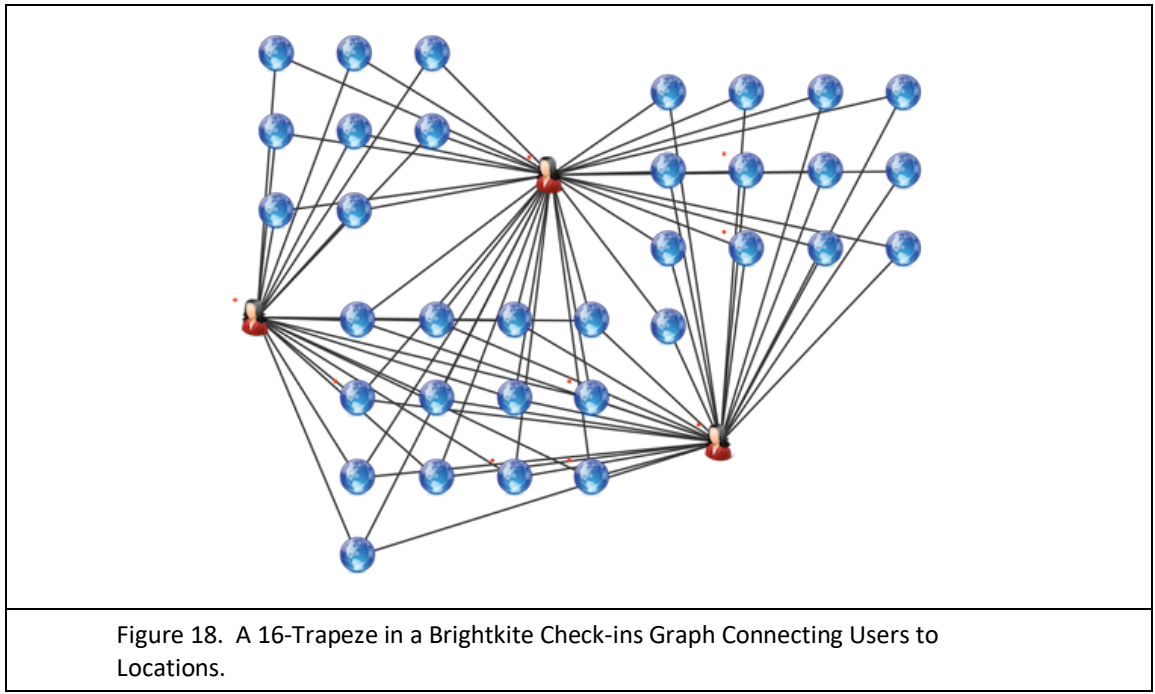

Figure 18.  A 16-Trapeze in a Brightkite Check-ins Graph Connecting Users to Locations.

Figure 19 shows three strong 8-trapezes in the (remaining) check-in graph sharing common locations.  Overlaid on the graph are dashed edges indicating friendships identified by the independent friendship graph.

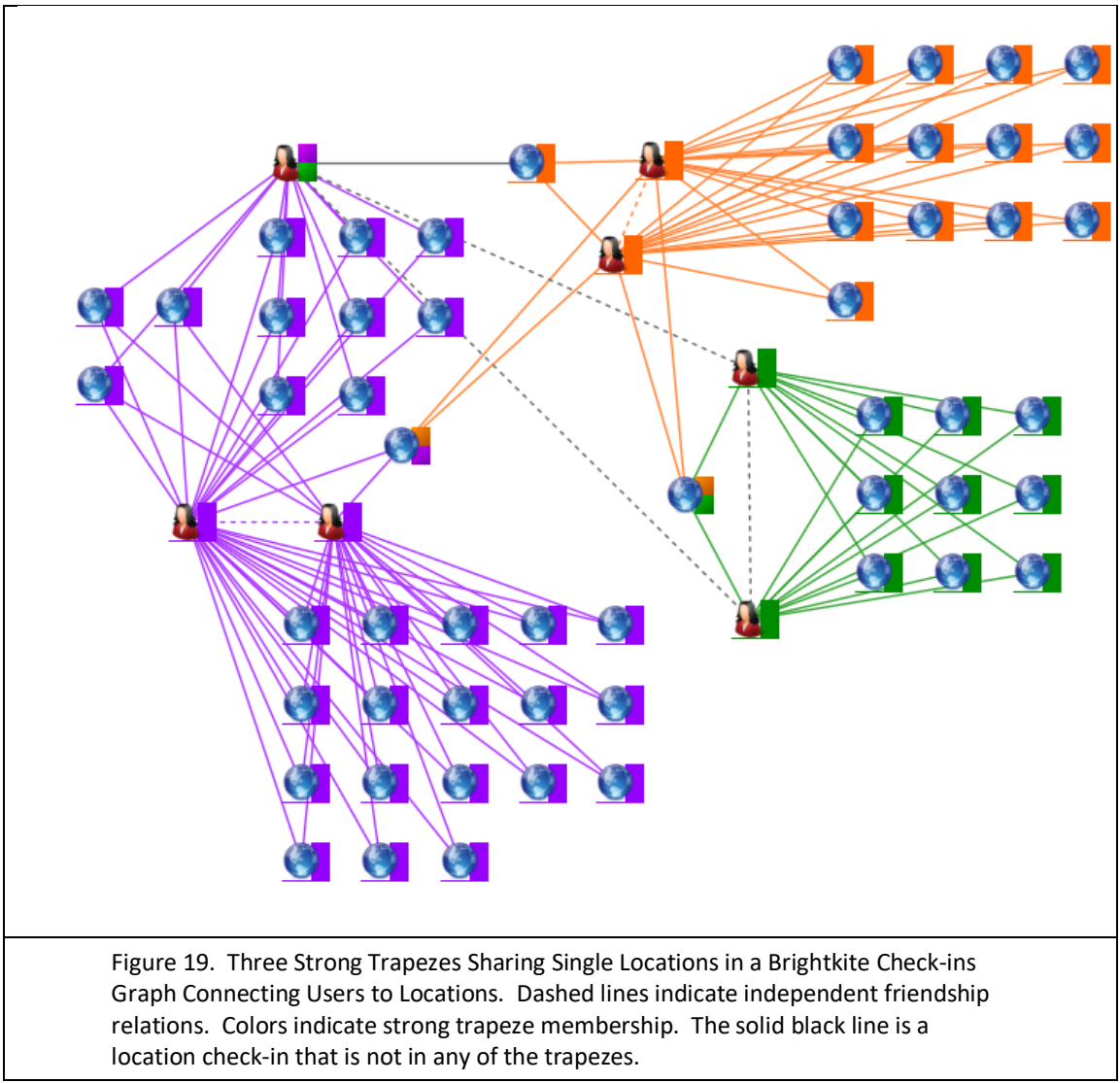

Figure 19. Three Strong Trapezes Sharing Single Locations in a Brightkite Check-ins Graph Connecting Users to Locations. Dashed lines indicate independent friendship relations. Colors indicate strong trapeze membership. The solid black line is a location check-in that is not in any of the trapezes.

Note that the time of check-ins was not considered. Experiments were run in which the locations were divided by date, so that two actors visiting the same location would be credited with coincidence only if their visits occurred on the same day. The result was a nearly-complete loss of predictive power of friendship (as indicated by the separate friendship graph) because of the sparsity of data.

# Alternative Computational Methods

Huang, *et al.* (2014) offer methods for finding all *k*-trusses in a graph that is too big to be contained in memory using a small number of scans of the graph from disk. In addition, they provide methods for updating trusses as edges are added or deleted in a dynamic environment. They also offer rapid querying of *k*-trusses containing any specified node, using an index system that scales linearly with the number of edges in the graph.

Looking at less-conventional computing platforms, Chen, *et al.* (2014) describe truss implementations for MapReduce (improving on those of Cohen 2009) and for a bulk synchronous parallel computation model such as Pragel (Grzegorz, *et al.*, 2010), in which computing nodes represent edges in the graph, and at each cycle computing nodes can receive information from their neighbors (who represent adjacent edges). Rossi (2014) describes fast, compact computation and general parallel implementation.

# Conclusion

Trusses and trapezes provide well-motivated definitions of communities, are easy to interpret, avoid problems of intertangled clusters, and are computationally efficient. They provide a sequence of nested subgraphs corresponding to levels of tightness that can be chosen globally or locally. Trusses have been used for anomaly detection in financial transaction data (Redmond, *et al.*, 2011), predicting the spread of disease (Malliaros, 2015), and network visualization (Zhang and Parthasarathy, 2012). With the introduction of the trapeze, these advantages can be brought to bear on multi-modal and bipartite graphs.

Many extensions and refinements of trusses and trapezes are useful: Strong trusses have long been used in practice for recognizing tighter subtrusses and identifying actors that bridge these tighter groups. Weighted versions offer the ability to have support also be a function of triangle (or rectangle) strength, accommodating the strength of constituent edges. Finally, summit trusses (and trapezes) permit the quick determination of the tightest structures across a graph of varying density while obviating the need for testing each value of support.

With their many advantages, the truss and trapeze are ideal for identifying social foci in network data.

# Acknowledgments

The author would like to thank Jessie Jamieson, Ian McCulloh, Kimberly Glasgow, Michael Castle, Alexander Perrone, and Anthony Johnson for their review of the manuscript and the improvements that followed, and for generous funding by the US Army Research Laboratory.